\documentclass{article}\usepackage[]{graphicx}\usepackage[]{color}
\makeatletter
\def\maxwidth{ %
  \ifdim\Gin@nat@width>\linewidth
    \linewidth
  \else
    \Gin@nat@width
  \fi
}
\makeatother

\definecolor{fgcolor}{rgb}{0.345, 0.345, 0.345}

\usepackage{framed}
\makeatletter
 {\par\unskip\endMakeFramed%
 \at@end@of@kframe}
\makeatother

\definecolor{shadecolor}{rgb}{.97, .97, .97}
\definecolor{messagecolor}{rgb}{0, 0, 0}
\definecolor{warningcolor}{rgb}{1, 0, 1}
\definecolor{errorcolor}{rgb}{1, 0, 0}
\usepackage{tabularx}
\usepackage{graphicx}
\usepackage{adjustbox}
\usepackage{changebar}
\usepackage{booktabs}
\usepackage{alltt}          
\usepackage{authblk}
\usepackage{fullpage}
\usepackage{amsthm}
\usepackage{graphicx}
\usepackage{url}
\usepackage{xcolor}
\usepackage{natbib}
\usepackage{amsmath}
\usepackage{amssymb}
\usepackage{arydshln}
\usepackage{hyperref}
\usepackage{algorithmicx}
\usepackage{algpseudocode}

\graphicspath{{./img/}}

\newtheorem{theorem}{Theorem}
\newtheorem{proposition}[theorem]{Proposition}%
\newtheorem{definition}{Definition}

\newtheorem{corollary}{Corollary}

\IfFileExists{upquote.sty}{\usepackage{upquote}}{}
\begin{document}

\title{A Bi-failure Mode Model for Competing Risk Modeling with HMC-Driven Bayesian Framework}

\author{Badamasi Abba$^1$\footnote{email: \texttt{badamasiabba@csu.edu.cn, badamasiabba@gmail.com}}\hskip 2em 
        Mustapha Muhammad$^2$\footnote{email: \texttt{mmmahmoud12@sci.just.edu.jo}}\hskip 2em 
        Muhammad Salihu Isa$^1$\footnote{email: \texttt{s-misa2@csu.edu.cn}}\hskip 2em 
        Jinbiao Wu$^1$\footnote{email: \texttt{wujinbiao@csu.edu.cn}}}
\affil{$^1$ School of Mathematics and Statistics, Central South University, Changsha, 410083, China \\ $^2$ Department of Mathematics, Guangdong University of Petrochemical Technology, Maoming, 525000, China}

\date{}

\maketitle

\begin{abstract}
   Bathtub failure rate (BFR) and roller-coaster failure rate (RCFR - a sequence of BFR and inverted BFR (IBFR)) shapes are among the non-monotone failure rate function (FRF) behaviors often observed in complex or competing risks (CR) datasets. Recent studies have introduced varied bathtub failure rate models for reliability modeling of such datasets. However, limited attention is paid to the reliability study of CR datasets characterized by RCFR. Motivated by this drawback, this paper proposes the so-called Bi-Failure Modes (BFM) model for robust reliability analysis of CR data exhibiting BFR, RCFR, and several other FRF shapes. The mean residual life function (MRLF) and cause-specific failure probabilities are studied in detail. The fundamental reciprocal relationships between the MRLF and FRF are established. We propose the Hamiltonian Monte Carlo (HMC)-based Bayesian framework for estimating the BFM parameters and its reliability attributes to offer greater computational efficiency and faster inference. Two CR datasets from electrode voltage endurance life and electrical appliance tests, respectively, characterized by BFR and RCFR behaviors, are employed to demonstrate the BFM adequacy. The recently introduced Bridge Criterion (BC) metric and other metrics are used to evaluate the BFM modeling performance against five recent methodologies under the maximum likelihood technique. The BFM compatibility with the two datasets is also examined. 
   
\end{abstract}


\section{ Introduction}\label{S1}
\,\indent Competing risk models are vital in reliability analysis, mainly in assessing the reliability of systems vulnerable to multiple causes of failure (COFs). The occurrence of different modes of failure can influence a system's overall performance and durability. Competing risks emerge when the failure of one component or mode may prevent the failure of others, demanding a broad insight into how varied causes interact. This is incredibly suitable in complex systems, where numerous components may fail due to different failure mechanisms.

Independent CR models suppose that the different failure processes operate independently, which means the failure times recorded from different causes are independently distributed random variables (see \cite{Samanta2019CRisk, Balakrishnan2008CRisk}). This hypothesis facilitates analysis and simplifies the computation of failure probabilities and times of failure for different risks without the complications caused by dependency. For example, in a mechanical system, one component may experience failure as a result of fatigue, while another component may experience failure owing to corrosion. Engineers may develop more precise predictions and maintenance procedures by independently modeling these failures. 

Contemporary improvements have underscored the significance of these models across diverse fields. For instance, researchers have employed CR models to study survival data where patients may succumb to various diseases or complications \citep{Austin2016}. \cite{HEISEY2009} assessed cause-specific mortality in ecological research that included CR. Besides, in engineering, CR models help assess the reliability of systems impacted by numerous COFs, hence facilitating the enhancement of design and maintenance strategies \citep{Meeker2022}. Ranjan and Upadhyay \citep{Ranjan2016} demonstrated the roles of CR model in the reliability modeling of vehicle failures.

\begin{figure}[ht]
	\centering	
	\includegraphics[width=3.75in, height=3.35in, keepaspectratio=false]{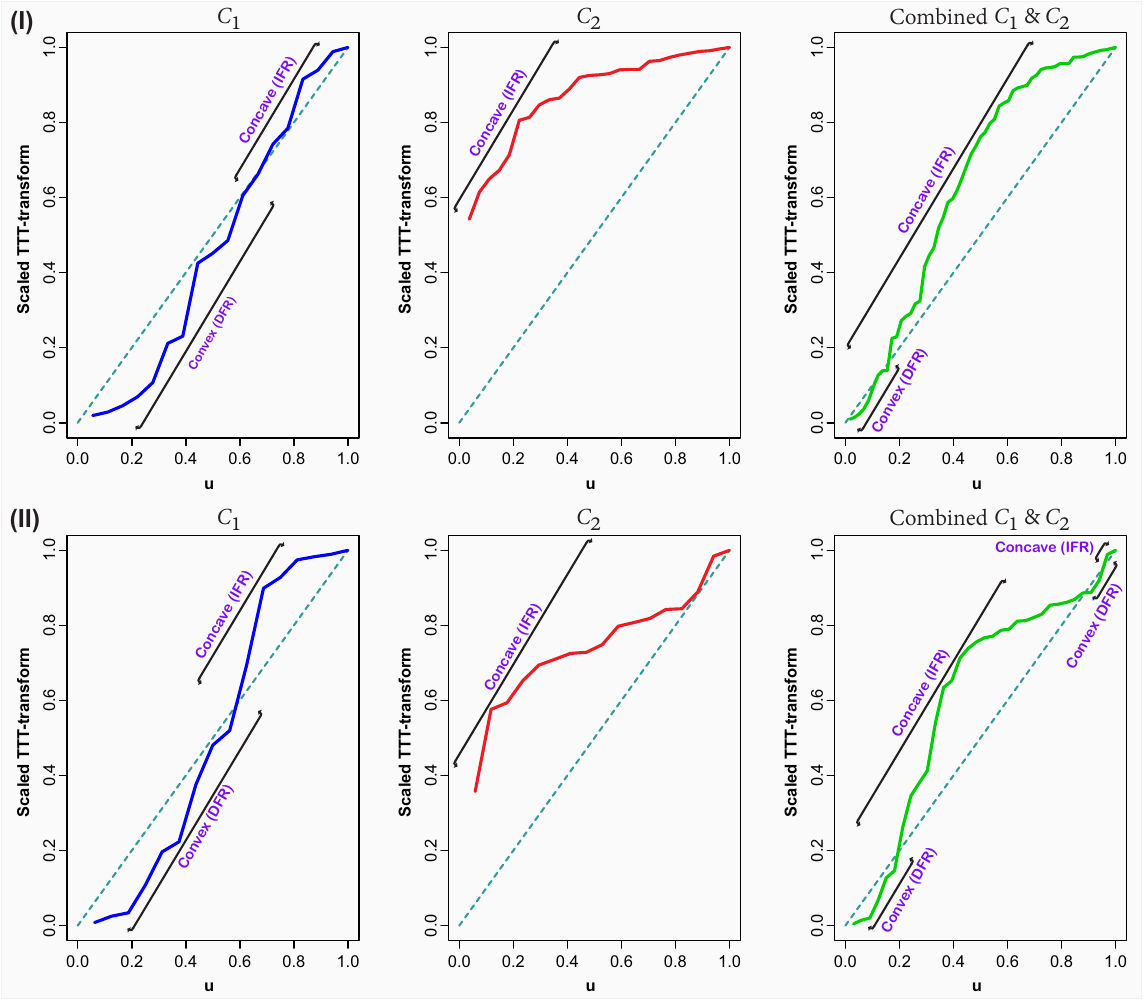}
	\centering
	\caption{\small Scaled TTT-transform plots for BFM datasets: (I) electrode times to failure portraying BFR, IFR and BFR for $C_{1}$, $C_{2}$ and combined $C_{1}$ \& $C_{2}$ and (II) electrical appliances failure and survival times depicting BFR, IFR and RCFR for $C_{1}$, $C_{2}$ and combined $C_{1}$ \& $C_{2}$, respectively.}
	\label{figTTT}
\end{figure}

The cause-specific and combined failure times (FTs) in CR or complex data studies are mostly found to exhibit different failure rate (FR) characteristics, including linear FR, such as the increasing and decreasing shapes, and nonlinear FR curves, such as BFR and RCFR curves, among others. To provide some highlights, \figurename{s \ref{figTTT}(I)} and \ref{figTTT}(II) depict the FR features through the TTT-transform graphs for two CR datasets. The first data describes the FTs of 58 electrodes subjected to a high-voltage stress test \citep{Doganaksoy2002}, and the second data set represents the FTs of 33 electrical appliances placed on a life test \citep{Lawless2003}. The two CR datasets were all taken to have two COFs (labeled as $C_{1}$ and $C_{2}$). \figurename{ \ref{figTTT}(I)} shows the TTT-transform plots for the cause-specific and combined first data set. It can be seen from the graphs that the FTs attributed to $C_{1}$, $C_{2}$ and combined ($C_{1}$ \& $C_{2}$) have convex-concave, concave, and convex-concave curves, which translate to BFR, increasing FR (IFR) and BFR curves, respectively. For the latter data set, the TTT-transform graphs in \figurename{\ref{figTTT}(II)} indicate convex-concave, concave, and  convex-concave-convex-concave curves corresponding to $C_{1}$, $C_{2}$ and combined ($C_{1}$ \& $C_{2}$). These translate to BFR, IFR, and RCFR shapes. 
One more recent illustration is obtained from \cite{Tang2015}, who studied 16 early failures of cable joints among the 31 registered failures by a power supply company between 2011 and 2014. The study recorded quality, installation, and unknown issues as three COFs. The aggregate data sets were demonstrated to have BFR (see \cite{Abba2022, Wang2022}).

These CR data characteristics have explored the relevance of incorporating FR feature information in CR data analysis. However, it was observed that different parametric reliability studies on these FT datasets, among many others, do not consider the cause-specific FR in the CR data modeling or were wrongly assumed to be described by some failure models. The studies by \cite{Tang2015} and \cite{Sarhan2010} are two obvious instances where CR models violates the cause-specific FR features of the data.

Besides that, the BFR is widely known for its roles in the three major facets of reliability assessments, that is, system reliability, hardware reliability, and human-related reliability, as recently elaborated in a review paper by \cite{Friederich2024}. Although our earlier examples of CR data are centered on hardware reliability, the BFR is also quite relevant in software and human-related reliability analysis (see \cite{Klutke2003,Friederich2024, Lazarova-Molnar2017}).

Numerous statistical failure models exhibiting diverse features were recently developed to handle CR or complex datasets characterized by two COFs. For instance, \cite{Singh2016} proposed and applied the so-called additive Perks-Weibull model to analyze the two COFs ($C_{1}$ \& $C_{2}$) electrode FT data by \cite{Doganaksoy2002}. The study assumed that Perks and Weibull models would each describe the FTs from one of the two COFs. The established model was shown to describe the BFR feature of the CR data. \cite{Shakhatreh2019} demonstrated the potential role of the log-normal modified Weibull model in modeling single and double competing risk FTs. It was portrayed not only as suitable for the reliability modeling of BFR but also for modified BFR and RCFR. \cite{Alhidairah2024} proposed a new CR model assuming independence between the COFs with blood cancer data application. The authors suppose that the FTs, due to two independent COFs, followed a generalized linear-exponential model and thus constructed the so-called additive generalized linear-exponential CR model. The study, however, does not dwell well on the FR characteristics of the proposed model. 
Other examples of CR models with BFR characteristics can be obtained in \cite{Al-essa2023, Mendez-Gonzalez2022a, Abba2023b,Thach2020,Mendez-Gonzalez2023}, and the references therein. Other types of CR models supposing varied CR scenarios are present in the literature. Two recent related instances, among others, are the partially observed modes of failure \citep{Chandra2023} and dependent competing risk processes \citep{Huang2024}.

For some failure mechanisms, the FR behavior of the combined FTs, in some instances, resembles more of the RCFR shape than the BFR. \cite{Wong1991RC} established that many electronic devices' FR features resemble RCFR curves, not BFR. He proposed several tenable physical explanations for how the RCFR shape formed to prove his assertion. Furthermore, \cite{Gupta2011RC} investigated the RCFR behavior of the extended generalized inverse Gaussian model. As earlier pointed out, the TTT-transform for the FTs of electrical appliances given in \figurename{\ref{figTTT}(II)} depicts a convex-concave-convex-concave shape, which translates to RCFR curve for the appliance data. However, this kind of FR complexity could not be adequately analyzed by most of the existing BFR or CR models, given that they are mostly constructed to handle the BFR curve, as can be seen in \cite{Alhidairah2024} and \cite{Al-essa2023} and the references therein. Nonetheless, the RCFR feature of this data set coincides with the proposed model FR curve shown in \figurename{ \ref{F1}(III)}. This suggests its suitability in the CR reliability modeling of FR characteristics of this form.

Bayesian methodology is an effective estimation procedure in statistical reliability analysis of diverse complex failure datasets. Since joint posterior distribution in CR modeling is frequently of intricate form, it is challenging to determine each model parameter's exact marginal posterior density and its associated reliability features. In tackling this complexity, multiple Markov chain Monte Carlo (MCMC) sampling procedures are considered to generate samples from the joint posterior distribution. The Metropolis-Hastings (MH) and other Gibbs algorithms are often utilized to perform the posterior analysis, which requires the marginal posterior density to either follow a known family of distribution or be log-concave. It is well-known that to achieve convergence to the target distribution, which is mostly in complicated form, the traditional MCMC sampling algorithms such as MH and Gibbs sampling may take a long time with less efficiency \citep{Zhang2022}. The popular MCMC methods are often infeasible in a small sample size case or high auto-correlation \citep{VandeSchoot2021}. The Hamiltonian Monte Carlo sampling algorithm (HMC-SA) method is one robust option, among others, to handle the small sample size and high auto-correlation problem properly. The HMC-SA employs a sequence of actions guided by first-order gradient information to circumvent the random walk of Metropolis-Hastings and sensitivity \citep{Saberzadeh2023, Betancourt2017}. Hence, in this article, we consider the HMC-SA for posterior sampling.

In addition to the competing risks and Bayesian estimation, censoring is also a frequent phenomenon in lifetime and reliability research. Among its different types, the Type-I and Type-II censoring schemes remain the most commonly encountered in different experiments \citep{Wang2018Censoring}. Although some literature favored the Type-II censoring scheme, numerous FTs data were recorded with Type-I censoring, particularly, the right censoring scheme \citep{Doganaksoy2002}.

In order to enhance the efficiency of CR analysis for two COFs problems while still accounting for the cause-specific and aggregate FR characteristics, the \textit{first contribution} of this paper is to develop a non-monotone failure rate (NMFR) model with adequacy to account for the cause-specific FR features in CR modeling. To achieve this, we present a Bi-Failure Modes (BFM) model - a hybridization of two baselines NMFR models, namely, the Dhillon \citep{Dhillon1980} and exponential power \citep{Smith1975} models. The \textit{second contribution} is to study the BFM reliability properties and evaluate its performance in CR data analysis. For better efficiency and faster inference of the proposed methodology, the \textit{third contribution} proposes estimating the model's parameters and its reliability characteristics using the HMC-SA-based Bayesian inference under the right censoring scheme.

This study is unique because, as far as we know, the study exploring the Dhillon and exponential power models as the baseline failure models for building a CR model is yet to be conducted. As noted earlier, these two models can analyze monotone failure rates and NMFRs, extending the robustness and applicability of the proposed CR model to a wide range of FR characteristics. Moreover, these specific baseline models are considered because the combined cause-specific FRs for some CR datasets possess NMFR features that differ from the commonly known BFR or IBFR curves. Besides constructing the BFM, we present and discuss several reliability properties of the model and investigate the fundamental reciprocal associations between the FRF and MRLF of the proposed model. The study employs the Hamiltonian Monte Carlo (HMC)-based Bayesian inference to ensure quick and efficient posterior analysis.
Additionally, the Bridge criterion (BC) metric \citep{Ding2017} and six other parametric and non-parametric performance metrics are deplored to evaluate the adequacy of the BFM against several other CR and non-CR models on two CR data sets. The BC metric is an information criterion constructed to hybridize the strengths of the Akaike information criterion (AIC) and Bayesian information criterion (BIC). It aims to achieve BIC's properties in the parametric analysis and AIC in the non-parametric setting \citep{Ding2017}.

The rest of the article is structured as follows: Section \ref{S2} introduce the BFM model and its reliability features. Description of the HMC-SA is given in Section \ref{S3}. Sections \ref{S4} and \ref{S5} present the Bayesian formulation and CR data illustration of the proposed model. The study is wrapped up in Section \ref{S6}.

\section{The Proposed BFM Model}\label{S2}

\,\;\indent This part describes the proposed BFM model, its basic functions and properties.  We investigate the fundamental reciprocal relationships between the MRLF and FRF. Two approaches for estimating the BFM's risks of failure are then derived and evaluated numerically.

Assume $c$ COFs are identified in a life-testing experiment and let these causes be represented by $C_{1}, \cdots, C_{c}$.  Suppose that for each cause $C_{k}$, the FT is described by the random variable $\mathcal{X}_{k},\, k=1, \cdots, c$ and is recorded alongside its associated COF. When all potential $c$ CoFs are in play, the observed FT of the system is the minimum of all FTs, denoted by $\mathcal{M}$. Hence, the random variable $\mathcal{X}=\min(\mathcal{X}_{1}, \cdots, \mathcal{X}_{c})$ defines the FTs of systems under investigation. Thus, the primary functions of $\mathcal{X}$ are given follows.

For for each cause $C_{k}$, there is corresponding reliability function (RF) denoted by $\bar{W}_{k}(x)$. The complete RF of the system, denoted by $\bar{W}(x)$, is defined as
\begin{eqnarray}
	\small	\begin{aligned}\label{S2E1}
		\bar{W}(x)&=Pr(\mathcal{X}>x)=Pr(\mathcal{X}_{1}>x, \cdots, \mathcal{X}_{c}>x)\\&=\prod_{k=1}^{c} {\bar{W}_{k}(x)}.
	\end{aligned}
\end{eqnarray}

The PDF, $w(x)$ can be obtained from \eqref{S2E1} as $w(x)=-\frac{\text{d}}{\text{d}x} \bar{W}(x)= -\frac{\text{d}}{\text{d}x}\left( \prod_{k=1}^{c} {\bar{W}_{k}(x)}\right) ,$ 
which represents the aggregate of probabilities of the $c$ different CoFs. Let for each cause $C_{k}$, the corresponding FRF be $r_{k}(x)$, with probability $r_{k}(x)\Delta x+o(\Delta x)$ that a working component or system will fail in the time interval ($x, \,\,x+\Delta x$). The overall FRF, denoted by $r(x)$ is the sum of $r_{k}(x)$ for $k=1, \cdots, c$, and can also be derived from \eqref{S2E1}, as given by
\begin{equation} \label{S2E2} 
	r(x)=-\frac{\text{d}}{\text{d}x} \log	\bar{W}(x) = \frac{w(x)}{\bar{W}(x)}   = \sum_{k=1}^{c} {r_{k}(x)}.
\end{equation} 
Define $I=k$ if system's failure occur due to cause $C_{k}$, for $k= 1, \cdots, c$. Then
\begin{equation*} \label{S1E4} 
	r_{k}(x)=\lim_{\Delta x\rightarrow0}\frac{Pr(\mathcal{X}<x+\Delta x, \, I=k|\mathcal{X}\ge x)}{\Delta x}.
\end{equation*} 

Now, suppose that the FTs attributed to $C_{1}$ is represent by a Dhillon random variable $\mathcal{X}_{1}$ and the FTs that occur due to  $C_{2}$ is described by an exponential power random variable $\mathcal{X}_{2}$. Thus, the PDFs and FRFs for $\mathcal{X}_{1}$ and $\mathcal{X}_{2}$, are accordingly, defined in Definition \ref{D1}--\ref{D2}, as follows:

\begin{definition}\label{D1}
	
	Suppose a random variable $\mathcal{X}_{1}$ follows Dhillon model \citep{Dhillon1980} having its scale and shape parameters, denoted by $\nu>0$ and $\theta>0$, respectively. Then the PDF and FRF of  $\mathcal{X}_{1}$ are presented as 
	\begin{equation} \label{S2E3} 
		w_{1}(x)=\frac{\theta\nu x^{\theta-1}}{\left( \nu x^{\theta} +1 \right) ^{2}},\,\, x>0,
	\end{equation} 
	and
	\begin{equation} \label{S2E4} 
		r_{1}(x)=\frac{\theta\nu x^{\theta-1}}{\nu x^{\theta} +1 },\,\, x>0.
	\end{equation} 
\end{definition}

Relating \eqref{S2E3} and \eqref{S2E4}, we have $\bar{W}_{1}(x)=\left( \nu x^{\theta} +1 \right) ^{-1}$ as the RF of $\mathcal{X}_{1}$. Where $\bar{W}_{1}(x)=	w_{1}(x)/r_{1}(x)=1-W_{1}(x)$. Dhillon \citep{Dhillon1980} establishes that the FRF of the model \eqref{S2E4} produces a decreasing curve when $\theta\le1$ and exhibits an IBFR for $\theta>1$. Despite the Dhillon flexibility, it has received little attention in reliability applications partly because it is not well-known in engineering and scientific communities. Thus, to promote its applicability, this study proposes to use the model as a component in conjunction with the exponential power model to build an FR-based CR model.

\begin{definition}\label{D2}
	For an exponential power random variable $\mathcal{X}_{2}$, the PDF and FRF of $\mathcal{X}_{2}$ are expressed as 
	\begin{equation} \label{S2E5} 
		w_{2}(x)=\tau\zeta  \left( \zeta x\right) ^{\tau-1}e^{\left( \zeta x\right) ^{\tau}}\exp{\left\lbrace 1-e^{\left( \zeta x\right) ^{\tau}} \right\rbrace   },\,\, x>0,
	\end{equation} 
	and 
	\begin{equation} \label{S2E6} 
		r_{2}(x)=\tau\zeta  \left( \zeta x\right) ^{\tau-1}e^{\left( \zeta x\right) ^{\tau}},\,\, x>0,
	\end{equation} 
	where $\tau>0$ and $\zeta>0$ are the continuously valued shape and scale parameters of the exponential power model \citep{Smith1975}.
\end{definition}
The RF is $\bar{W}_{2}(x)=\exp{\left\lbrace 1-e^{\left( \zeta t\right) ^{\tau}} \right\rbrace}.$ \cite{Smith1975} constructed the exponential power as life-testing model and is relevant in an instance when FTs  can be characterized by U-shaped bathtub or increasing FR. Numerous extensions of exponential power for modeling varied reliability FTs have existed in the literature (see, for instance \cite{Korkmaz2021,Abba2023a}).

Suppose $\mathcal{X}_{1}$ and $\mathcal{X}_{2}$ are independently, identically distributed. Thus, letting $\mathcal{X}=\min(\mathcal{X}_{1}, \mathcal{X}_{2})$ to define the FTs of systems under investigation, we built the primary functions of the minimum of Dhillon and exponential power random variables as given in \eqref{S2E7}-\eqref{S2E9}. That is, by following \eqref{S2E1}--\eqref{S2E6}, we have

\begin{equation} \label{S2E7} 
	\small\begin{aligned}
		w_{\text{BFM}}\left(x, \boldsymbol{\varpi}\right)=\frac{\left(\frac{\nu\theta x^{\theta-1}}{ (\nu x^{\theta} +1)^{2}  }+\frac{\tau\zeta  \left( \zeta x\right) ^{\tau-1}e^{\left( \zeta x\right) ^{\tau}}}{\nu x^{\theta} +1} \right) }{\exp{\left\lbrace e^{\left( \zeta x\right) ^{\tau}} -1\right\rbrace  }}, \,\,\, x>0,
	\end{aligned}
\end{equation} 
where $\boldsymbol{\varpi}=(\nu, \tau, \theta,  \zeta)^{\prime}$ is a continuously vector of valued parameters having $\tau, \theta >0$ as shape parameters and $\nu, \zeta\ge0$ representing the scale parameters of the model. Consequently, we write $\mathcal{X}\sim$BFM$(\boldsymbol{\varpi})$. The BFM model \eqref{S2E7} is built with the assumption that the FTs for each of the two competing CoFs follow a certain NMFR model. That is, the failures from one of the CoFs has Dhillon random variable characteristic while the failures due to the other CoF is assumed to have an exponential-power model. 

The RF and FRF of Equation \eqref{S2E7} are determined by the appropriate substitutions of \eqref{S2E3}-\eqref{S2E6} into \eqref{S2E1} and \eqref{S2E2}. Therefore, we have
\begin{equation} \label{S2E8} 
	\bar{W}_{\text{BFM}}\left(x, \boldsymbol{\varpi}\right)=\frac{\exp{\left\lbrace 1-e^{\left( \zeta x\right) ^{\tau}} \right\rbrace  } }{\nu x^{\theta} +1}, \,\,\, x>0
\end{equation} 
and
\begin{equation} \label{S2E9} 
	r_{\text{BFM}}\left(x, \boldsymbol{\varpi}\right)=\frac{\theta\nu x^{\theta-1}}{ \nu x^{\theta} +1 }+\frac{\tau\zeta \left( \zeta x\right) ^{\tau-1}}{e^{-\left( \zeta x\right) ^{\tau}}},\quad x>0. 
\end{equation} 
The RF \eqref{S2E8} can be observed to have exhibited an easygoing structure, and hence facilitates the computations and inference of the BFM parameters alongside its reliability properties. We have graphically demonstrated in \figurename{s \ref{F1}}(I), (III) and (V), that the FRF in \eqref{S2E9} can described BFR, IBBFR (combination of inverted Bathtub and BFR), and RCFR behaviors, among several others depending on the BFM shape parameters $\tau$ and $\theta$. 
The CDF, ${W}_{\text{BFM}}\left(x, \boldsymbol{\varpi}\right)$ of $\mathcal{X}$ can be derived from \eqref{S2E8} via the relation $W_{\text{BFM}}\left(x, \boldsymbol{\varpi}\right)=1-\bar{W}_{\text{BFM}}\left(x, \boldsymbol{\varpi}\right)$. The BFM cumulative failure rate function (CFRF), denoted by $	R_{\text{BFM}}\left(x, \boldsymbol{\varpi}\right)$, is defined as
\begin{equation} \label{S2E10} 
	R_{\text{BFM}}\left(x, \boldsymbol{\varpi}\right)=\int_{0}^{x}r(m)\text{d}m= e^{\left( \zeta x\right) ^{\tau}} +\ln(\nu x^{\theta} +1)-1.
\end{equation} 


\subsection{The MRLF of BFM Model}

\,\indent The mean remaining life function, also known as mean residual life function (MRLF), is a crucial notion in reliability analysis and survival studies, describing the average remaining life of a system provided it has survived up to a definite time, say $\tilde{x}$. The MRLF as a reliability metric offers vital comprehension into the durability and degradation of systems, which is essential for decision-making in maintenance, system design, and warranty services. Understanding MRLF empowers analysts, particularly engineers, to evaluate the performance and reliability of components over time, expediting better predictions about when a component may fail and hence enhancing maintenance schemes.
The MRLF concept is rooted in survival analysis, where it is mathematically defined as the expected remaining life of a subject at time $\tilde{x}$, expressed as  $\mu(\tilde{x})=\mathbb{E}(\mathcal{X}-\tilde{x} | \mathcal{X} > \tilde{x})=\int_{0}^{\infty}\bar{W}(x+\tilde{x})\text{d}x/\bar{W}(\tilde{x}),$
where $\mathcal{X}$ denotes the lifetime of the system or subject.

Let $\mu_{\text{BFM}}(\tilde{x})$ be the MRLF of $\mathcal{X}$ that follows the BFM model. By applying algebraic operations, it can be deduced that the MRLF of $\mathcal{X}$ at time point $\tilde{x}$ corresponds to the expression stated in Proposition \ref{th1}.

\begin{proposition}\label{th1}
	Let $\mathcal{X}\sim\text{BFM}(\boldsymbol{\varpi})$. The MRLF of $\mathcal{X}$ is
	\begin{eqnarray*}\label{SMRL}
		\small\begin{aligned}
			\mu_{\text{BFM}}(\tilde{x})=\frac{e}{ \bar{W}(\tilde{x})}\sum_{\ell\ge0}{ \frac{{(-\nu)^{\ell}}}{\tau\zeta^{\ell\theta+1}}\mathcal{I}(\tilde{x}, \ell) },
		\end{aligned}
	\end{eqnarray*}
	where $\mathcal{I}(\tilde{x}, \ell)=\int_{e^{(\zeta\tilde{x})^{\tau}}}^{\infty}{\left(\ln y \right)^{\frac{\ell\theta+1}{\tau}-1}y^{-1}e^{-y}\text{d}y .}$ 
\end{proposition}
\textbf{Proof.} The proof for Proposition \ref{th1} is derived by appropriately substituting \eqref{S2E8} into $\mu(\tilde{x})=\int_{0}^{\infty}\bar{W}(x+\tilde{x})\text{d}x/\bar{W}(\tilde{x})$, letting $\left(\nu(x+\tilde{x})^{\theta} \right) ^{-1}=\sum_{\ell\ge0}(-\nu)^{\ell}(x+\tilde{x})^{\ell\theta}$ and $y=e^{\left( \zeta(x+\tilde{x})^{\tau} \right) }$. For $e^{(\zeta\tilde{x})^{\tau}}=1$, the integral $\mathcal{I}(\tilde{x}, \ell)$ will produce $\Gamma\left(\frac{\ell\theta+1}{\tau}\right)E^{\frac{\ell\theta+1}{\tau}-1}_{1}(1)$ by applying the generalized integro-exponential function \citep{Milgram1985}.

The MRLF and FRF are two fundamental concepts in reliability modeling, and they are inherently interconnected in describing the behavior of life distributions of systems over time. Comprehending the connection between them is essential for proper reliability assessment, maintenance strategy, and formulating well-informed engineering decisions.
The following relationship intimately links the MRLF and FRF:
\begin{equation*}
	\mu^{\prime}(\tilde{x})=\mu(\tilde{x})r(\tilde{x})-1.
\end{equation*}
This suggests
\begin{equation}\label{E17_MRL}
	r(\tilde{x})=\dfrac{\mu^{\prime}(\tilde{x})+1}{\mu(\tilde{x})},
\end{equation}
where $\mu^{\prime}(\tilde{x})=\text{d}\mu(\tilde{x})/\text{d}x$. This relationship means that the two reliability functions $\mu(\tilde{x})$ and $r(\tilde{x})$ can autonomously determine the  distribution of failure time distribution of $\mathcal{X}$. Equation \eqref{E17_MRL} establishes that the reciprocals of MRLF and FRF of $\mathcal{X}$ are equivalent as $\mu^{\prime}(\tilde{x})/\mu(\tilde{x})$ approaches zero.

The MRLF is admitted to provide more precise information than the FRF in maintenance and replacement schemes. The FRF gives information on component or system failure inside a specific time window after a given time $\tilde{x}$. In contrast, the MRLF indicates the time to failure beyond $\tilde{x}$. Multiple studies, such as those by \cite{Shakhatreh2019} and \cite{Abba2023b}, have examined the link between the FRF and MRLF by analyzing their forms and change points. \cite{Gupta1995} establish that if $\mu>1/ r(0)$ and the system's FTs exhibits a BFR, then the corresponding MRL will display an inverted bathtub (IB) pattern. Similarly, if $\mu<1/ r(0)$ and the system's FTs depicts a IBFR, then the corresponding MRL will display a bathtub curve pattern. The authors argue that it is nearly unattainable to obtain these relationships using conventional analytical approaches since the FRF and MRLF are so intricate in most models. For a BFR with a single change point at $0\le x_{r}<\infty$, the MRLF will exhibit an IB shape with one turning point at $x_{\mu}$, where $x_{\mu} \le x_{r}$, which mean that the change point of the MRLF occurs before the turning point of its associated FRF. 

We can apply the MRLF to determine the optimal burn-in period ($b_{opt}$) of a product lifetime that exhibits a BFR. \cite{Mi1995} establishes that the an appropriate choice of $b_{opt}$ can be by maximizing the MRLF, such that $\mu^{\ast}=\mu(b_{opt}) = \max_{x\ge0}\mu(x)$, where $\mu^{\ast}$ symbolizes the MRLF's change point. It is obtained by differentiating $\mu(x)$ with respect to $\mathcal{X}$, with $b_{opt}$ as the value that maximizes $\mu(x)$. Other procedures for burn-in estimations can be found in \cite{Cha2012} and \cite{Jiang2008}, and the references therein.

\begin{figure}[h!]
	\centering	
	\includegraphics[width=3.15in, height=4.7in, keepaspectratio=false]{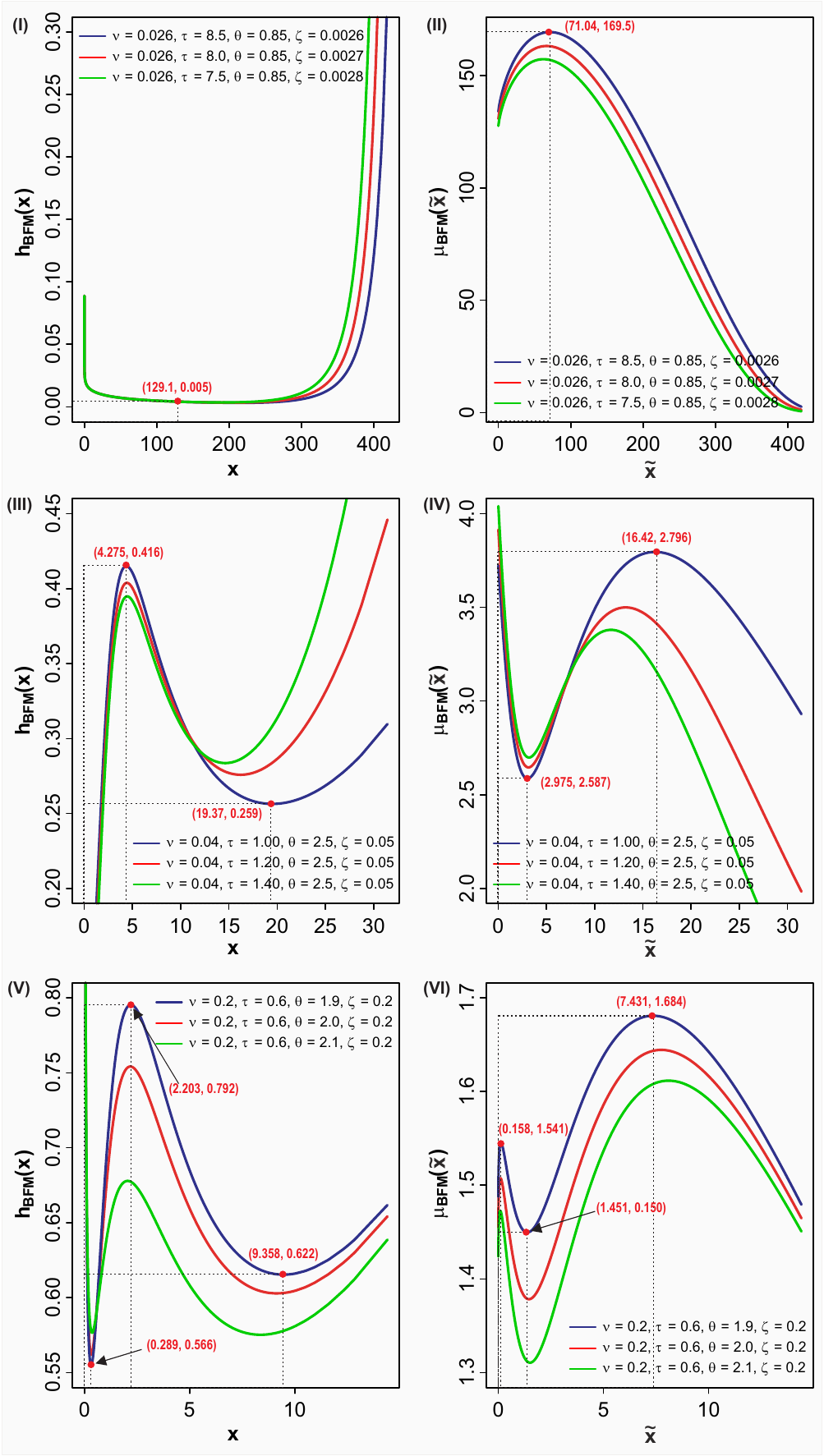}
	\centering
	\caption{Graphs of FRF and MRLF of the proposed BFM model: (I)-(II) BFR and IB-MRL for $\tau>1, \theta<1$, (III)-(IV) IBBFR and BIB-MRL for $\tau,\,\theta\ge1$, and (V)-(VI) RCFR and IRC-MRL for $\tau<1, \theta>1$.}
	\label{F1}
\end{figure}

Interestingly, the effect of the reciprocal relationship between the FRF and MRLF of the proposed BFM model can be clearly observed in \figurename{ \ref{F1}}. These plots depict the patterns of the FRF curves and their associated MRLF curves. \figurename{ \ref{F1}(I)} showcases that the FRF has a BFR shape with a long-constant portion, and its turning point occurred at $x_{r}=129.1$. \figurename{ \ref{F1}(II)} shows the corresponding behavior of the MRLF, which depicts an IB curve with a turning point at $x_{\mu}=71.04$. \figurename{s \ref{F1}}(III)-(IV) illustrates that the FRF exhibits an IBBFR pattern (with two change points at $x_{r}=4.275$ and 19.37). Meanwhile, its corresponding MRLF depicts a bathtub-inverted bathtub (BIB) curve with turning points at $x_{\mu}=2.975$ and 16.42, respectively. \figurename{s \ref{F1}(V)-(VI)} demonstrate RCFR and inverted-RC (IRC) curves, each with three change points ($x_{r}=0.289, 2.203, 9.358,$ and $x_{\mu}=0.158, 1.451,$ and 7.431), for the FRF and MRLF, respectively. Indeed, all the graphs have demonstratively portrayed the reciprocal relationship between the FRF and MRLF and that the turning points of the MRLF occur before the change points of the FRF. That is, $x_{r} \ge x_{\mu}$. 

\subsection{The MTTF of BFM Model}
\,\indent The MTTF measures the average period between two instances of failure. Understanding the duration of software performance before failure is a crucial measure that enables developers to anticipate and proactively address challenges, enhancing their ability to foresee failures and stay ahead of potential problems. At $\tilde{x}=0$,  $\mu_{\text{BFM}}(0)=\int_{0}^{\infty}{\bar{W}(x)\text{d}x=\text{MTTF}}$, and thus, we have the subsequent corollary for the MTTF following proposition \ref{th1}.
\begin{corollary}\label{th2}
	Let $\mathcal{X}\sim\text{BFM}(\boldsymbol{\varpi})$. The MTTF of $\mathcal{X}$, denoted by $	\text{MTTF}_{\text{BFM}}$, is
	\begin{eqnarray*}\label{MTTF}
		\small\begin{aligned}
			\text{MTTF}_{\text{BFM}}=\frac{e}{ \tau}\sum_{\ell\ge0}{\left[ \frac{{(-\nu)^{\ell}}}{\zeta^{\ell\theta+1}}\Gamma\left(\frac{\ell\theta+1}{\tau}\right)E^{\frac{\ell\theta+1}{\tau}-1}_{1}(1) \right]}.
		\end{aligned}
	\end{eqnarray*}
\end{corollary}

\subsection{Risks Analysis}	
\,\indent The probability of a system failing due to cause $C_{k}$ is described by $F_{k}(x)=F_{k}=Pr(\mathcal{X}\le x, I=k), k=1, \cdots, c.$ These probabilities are called cause-specific probabilities (risks) of failure or incidence functions and are expressed by
\begin{equation}\label{IF}
	F_{k}(x)=\int_{0}^{x}{r_{k}(t)\bar{W}(t)\text{d}t},
\end{equation}
Note that a system or individual must survive up to time $t$, and its failure at time $t$ attributed to cause $C_{k}$ is not connected to other CoFs; thus, led to the appearance of the term  $r_{k}(x)\bar{W}(x)$ in \eqref{IF}. Hence, for the proposed BFM model with only two CoFs $C_{k},\, k=1, \,2,$ the probabilities of failure due to $C_{1}$ and $C_{2}$ following Dhillon and exponential power models, accordingly, are derived using \eqref{IF} and provided in Propositions \ref{IF1}- \ref{IF2}.
\begin{proposition}\label{IF1}
	Given the Dhillon FRF, $r_{1}(x, \nu, \theta)=\frac{\theta\nu x^{\theta-1}}{ \nu x^{\theta} +1 }$ and the RF of the BFM model provided in \eqref{S2E8}, the risk of failure attributed to $C_{1}$ in the presence of $C_{2}$ is defined as
	\begin{eqnarray*}\label{IFE1}
		\small		\begin{aligned}
			F_{1}=	F_{1}(\infty)=\frac{\nu\theta e}{\tau}\sum_{\ell\ge0}{ \frac{{(\ell+1)(-\nu)^{\ell}}}{\zeta^{{\theta(\ell+1)}}}\Gamma\left(\frac{\theta(\ell+1)}{\tau}\right)E^{\frac{\theta(\ell+1)}{\tau}-1}_{1}(1)}. 
		\end{aligned}
	\end{eqnarray*}
\end{proposition}
\begin{proposition}\label{IF2}
	For an exponential power FRF, $r_{2}(x, \tau, \zeta)=\tau\zeta  x^{\tau-1}e^{\zeta x^{\tau}}$ and the RF of the BFM model expressed in \eqref{S2E8}, the probability of failure from $C_{2}$ in the presence of $C_{1}$ is defined as
	\begin{eqnarray*}\label{IFE2}
		\small	\begin{aligned}
			F_{2}=e\sum_{\ell,s\ge0}{ \frac{{(-\nu)^{\ell}}}{s!\zeta^{\ell\theta}}\Gamma\left(\frac{\ell\theta+\tau(s+1)}{\tau}\right)E^{\frac{\ell\theta+\tau(s+1)}{\tau}-1}_{1}(1)}. 
		\end{aligned}
	\end{eqnarray*}
\end{proposition}
\textbf{Proof.} To prove Propositions \ref{IF1} and \ref{IF2}, we apply the appropriate $r_{k}(x),\, k=1,\,2,$ and $\bar{W}(x)$ into \eqref{IF} and then perform some algebra steps to finish the proofs, as described under the proof of Proposition \ref{th1}.

As an alternate systematic explanation of the probability of failure from defined causes in the presence of competing risks under the so-called \textit{proportionality assumption}. 
In the case of $c$ causes, the chance of failure from cause $C_{k}$, where $k=1, \cdots, c$, is precisely defined as
\begin{equation*}
	F_{k}(x)=\frac{r_{k}(x)}{r(x)},
\end{equation*}
independent of $x$ for each $k$. Thus, for $k=1$ and $2$, the probabilities of failure for the Dhillon and exponential power related causes are given as
\begin{eqnarray}\label{IF_Chiang1}
	F_{1}(x)=\frac{\theta\nu x^{\theta-1}e^{-\zeta x^{\tau}}}{\theta\nu x^{\theta-1}e^{-\zeta x^{\tau}}+\tau\zeta  x^{\tau-1}(\nu x^{\theta} +1 )} 
\end{eqnarray}
and 
\begin{eqnarray}\label{IF_Chiang2}
	F_{2}(x)=\frac{\tau\zeta  x^{\tau-1}(\nu x^{\theta} +1 )}{\theta\nu x^{\theta-1}e^{-\zeta x^{\tau}}+\tau\zeta  x^{\tau-1}(\nu x^{\theta} +1 )}.
\end{eqnarray}

Similarly, the probabilities of failures can also be calculated by appropriately substituting $r_{k}(x)$ and $\bar{W}(x)$ in \eqref{IF} and then apply the one-dimensional integral function \verb|integrate()| in \verb|R| software. Therefore, we subsequently evaluate the proposed probabilities of failure formula with reference to the one-dimensional integral approach.

\tablename{ \ref{T1}} provides several estimates of the risks of failure, $	\hat{F}_{1}$ and $\hat{F}_{2}$ from causes $C_{1}$ and $C_{2}$. The values were computed using the numerical integration function (termed as $\mathcal{P}1$), proportionality assumption method (given by equations \eqref{IF_Chiang1}-\eqref{IF_Chiang2}, $\mathcal{P}2$) and propositions \ref{IF1}- \ref{IF2} ($\mathcal{P}3$), respectively. We noted that, among all the arbitrary chosen parameter values, the estimates from $\mathcal{P}1$ and $\mathcal{P}3$ are equal and are sum up to unity (i.e., $\hat{F}_{1}+\hat{F}_{2}=1$), except when the chosen values for $\tau$ and $\theta$ are above one. In these instances, $\hat{F}_{1}$ and $\hat{F}_{2}$ are intimately close but not exactly the same. It is noticed from \tablename{ \ref{T1}} that, the estimates derived from $\mathcal{P}2$ are somewhat closed to that of $\mathcal{P}1$ only when $\tau<1, \theta>1$ or $\tau>1, \theta<1$. Furthermore, we noted that few failures are attributed to $C_{1}$ ($\hat{F}_{1}=1.58\%$) while  $98.42\%$ of the failures occurred due to $C_{2}$ for $\varpi=(0.01,\, 0.6,\, 2.0,\, 0.6)$ under $\mathcal{P}1$, $\mathcal{P}2$ and $\mathcal{P}3$.  Evaluating the derived probabilities of failures under the different settings of the BFM shape parameters, the findings reveal that over 92\% of failures in all the scenarios resulted from $C_{2}$ except when $\tau<1$ and $\theta<1$. In this case, cause $C_{1}$ have recorded about 3\% (using $\mathcal{P}2$) and 30\% (under $\mathcal{P}3$) of the failures.  Therefore, the results pointed out the suitability of $\mathcal{P}3$ over $\mathcal{P}2$ for calculating the probability of failures under varied BFM competing risk problems.  

\begin{table}[h]
	\caption{\small Various illustrations of BFM  probabilities of failure for the causes $C_{1}$ and $C_{2}$ based on 	$\mathcal{	P}1$, 	$\mathcal{	P}2$ and 	$\mathcal{	P}3$ at some chosen values of $\boldsymbol{\varpi}=(\nu, \tau, \theta, \zeta)$.}
	\label{T1}
	\centering
	\begin{adjustbox}{width=3.3in}
		\begin{tabular}{lcccc}\toprule
			&($\nu, \tau, \theta, \zeta$)& \multicolumn{2}{c}{Risk of failure }&Overall\\ \cline{3-4}
			Method &$\downarrow$&$\hat{F}_{1}$ &$\hat{F}_{2}$&$\downarrow$ \\
			\midrule
			$\mathcal{	P}1$&$(0.01,\, 0.6,\, 2.0,\, 0.6)$ &0.0158&0.9842&1.0000\\	
			$\mathcal{	P}2$&&0.0159&0.9841&1.0000\\
			$\mathcal{	P}3$&&0.0158&0.9842&1.0000\\
			\cline{1-5}
			$\mathcal{	P}1$	&$(0.05,\, 0.7,\, 6.0,\,2.8)$&0.0005&0.9995&1.0000\\
			$\mathcal{	P}2$&&0.0083&0.9917&1.0000\\
			$\mathcal{	P}3$	&&0.0005&0.9995&1.0000\\\cline{1-5}
			$\mathcal{	P}1$	&$(0.01,\, 1.5,\, 0.3,\, 0.6)$&0.0098&0.9902&1.0000\\
			$\mathcal{	P}2$&&0.0044&0.9956&1.0000\\
			$\mathcal{	P}3$	&&0.0098&0.9902&1.0000\\\cline{1-5}
			$\mathcal{	P}1$	&$(0.5,\, 0.25,\, 0.05,\, 0.8)$&0.2991&0.7009&1.0000\\
			$\mathcal{	P}2$&&0.0302&0.9698&1.0000\\
			$\mathcal{	P}3$&&0.2990&0.7009&0.9999\\\cline{1-5}
			$\mathcal{	P}1$	&$(0.5,\, 3.0,\, 8.0,\, 1.2)$&0.0533&0.9467&1.0000\\
			$\mathcal{	P}2$&&0.0776&0.9224&1.0000\\
			$\mathcal{	P}3$	&&0.0548&0.9456&1.0004\\ 
			\bottomrule	
		\end{tabular}
	\end{adjustbox}
\end{table}

\section{The HMC-SA}\label{S3}
\,\indent This part provides a brief discussion of the HMC-SA. A detailed discussion of the algorithm can be found in \cite{Betancourt2017,Ranjan2021,Abba2023b,Muhammad2025}, and the references therein. One can also refer to \cite{Thomas2021} for an \verb|R| software implementation of the HMC-SA in a more digestible form for practitioners. 

The HMC-SA is expected to improve efficiency due to its utilization of a proposal based on the information gained from the posterior itself. Let $\Omega(\boldsymbol{\varpi}|\mathbf{x})$ be a joint posterior distribution from which we propose to generate a random sample. We define a set of $g$ ancillary variables $\kappa_{1}, \cdotp, \kappa_{g}$ which are assumed to be described by Gaussian distribution $N_{g}(0, q_{p})$, where $p=1, \cdots, g$. We express the joint density of $\varpi_{1},\cdots, \varpi_{g}$ and $\kappa_{1}, \cdots, \kappa_{g}$, as
\begin{equation}\label{S3E1}
	\exp\left\lbrace\ln\left(\Omega(\varpi_{1},\cdots, \varpi_{g}|\mathbf{x}) \right) -\sum_{p=1}^{g}{\frac{\kappa_{p}^{2}}{2q_{p}}} \right\rbrace.
\end{equation} 
Let $\boldsymbol{\kappa}$ denotes the ancillary variables' vector and suppose $H(\boldsymbol{\varpi}, \boldsymbol{\kappa})$ represents the negative logarithmic form of the joint density in \eqref{S3E1}. This translates to
\begin{equation}\label{S3E2}
	\begin{aligned}
		H(\boldsymbol{\varpi}, \boldsymbol{\kappa})&= -\ln\left(\Omega(\boldsymbol{\varpi}|\mathbf{x})\right) +\frac{\boldsymbol{\kappa}^{\prime}\boldsymbol{Q}^{-1}\boldsymbol{\kappa}}{2},
	\end{aligned}
\end{equation}
where $\frac{\boldsymbol{\kappa}^{\prime}\boldsymbol{Q}^{-1}\boldsymbol{\kappa}}{2}=\sum_{p=1}^{g}{\frac{\kappa_{p}^{2}}{2q_{p}}}$, $\boldsymbol{Q}$ represents the diagonal matrix with entries $q_{1}, \cdots, q_{g}$. Equation \eqref{S3E2} describes the Hamiltonian where the first term explains the potential energy at a specific position $\boldsymbol{\varpi}$ and the second term describes the kinetic energy with the vector of momentum $\boldsymbol{\kappa}$. Observing the partial derivatives of \eqref{S3E2} for $\boldsymbol{\varpi}$ and $\boldsymbol{\kappa}$ led us to the Hamiltonian equations of motion, given as
\begin{align}
	&\boldsymbol{\bar{\varpi}}=	\frac{\partial H(\boldsymbol{\varpi}, \boldsymbol{\kappa})}{\partial \boldsymbol{\kappa}}= \boldsymbol{Q}^{-1}\boldsymbol{\kappa},\label{S3E3} 
	\\
	&	\boldsymbol{\bar{\kappa}}=	-\frac{\partial H(\boldsymbol{\varpi}, \boldsymbol{\kappa})}{\partial \boldsymbol{\varpi}}= \Delta_{\boldsymbol{\varpi}}\ln\left(\Omega(\boldsymbol{\varpi}|\mathbf{x})\right),\label{S3E4} 
\end{align} 
where $\boldsymbol{\bar{\varpi}}$ and $\boldsymbol{\bar{\kappa}}$ in \eqref{S3E3} and \eqref{S3E4} represent the partial derivatives of the position and momentum vectors, respectively, with respect to time. Note that we used $\Delta_{\boldsymbol{\varpi}}$ in \eqref{S3E4} to denotes the gradient of $\ln\left(\Omega(\boldsymbol{\varpi}|\mathbf{x})\right)$.

Although there are several approaches to solving the Hamiltonian equations (see \cite{Neal2011}), many considerations are on the Leapfrog methodology. This approach enables the discretization of the Hamiltonian dynamics for a number of preset step sizes ($\epsilon>0$) with steps as follows:
\begin{align}\label{S3E5}
	&\kappa^{(\alpha+\epsilon/2)}	=\kappa^{\alpha}+\frac{\epsilon}{2}\Delta_{\boldsymbol{\varpi}}\ln\left(\Omega(\boldsymbol{\varpi}^{\alpha}|\mathbf{x}) \right),\nonumber\\
	&	\varpi^{(\alpha+\epsilon)}	=\varpi^{\alpha}+\epsilon\boldsymbol{Q}^{-1}\boldsymbol{\kappa}^{(\alpha+\epsilon/2)},\\
	&	\kappa^{(\alpha+\epsilon)}	=	\kappa^{(\alpha+\epsilon/2)}+\frac{\epsilon}{2}\Delta_{\boldsymbol{\varpi}}\ln\left(\Omega(\boldsymbol{\varpi}^{(\alpha+\epsilon)}|\mathbf{x}) \right),\nonumber
\end{align}
where the iteration step is denoted by $\alpha$. Iterating the system of equations in \eqref{S3E5} to $L$ Leapfrog number of steps produces a proposal point $\left(\boldsymbol{\varpi}^{L}, \boldsymbol{\kappa}^{L}\right)$. Immediately we obtained $\left(\boldsymbol{\varpi}^{L}, \boldsymbol{\kappa}^{L}\right)$, we then proceed to the step of Metropolis acceptance with a probability of acceptance expressed as
\begin{align}\label{S3E6}
	\eta=\min(1,\phi(\boldsymbol{\varpi}, \boldsymbol{\varpi}^{L}, \boldsymbol{\kappa}, \boldsymbol{\kappa}^{L})),\end{align}
where $\phi(\boldsymbol{\varpi}, \boldsymbol{\varpi}^{L}, \boldsymbol{\kappa}, \boldsymbol{\kappa}^{L})=\exp\left\lbrace {H(\boldsymbol{\varpi}, \boldsymbol{\kappa})}-{H(\boldsymbol{\varpi}^{L}, \boldsymbol{\kappa}^{L})}\right\rbrace.$ The acceptance probability is therefore used to derive a sample from the posterior distribution. Initial values of $\left(\boldsymbol{\varpi}, \boldsymbol{\kappa}\right)$ are needed to iteratively generate the proposal point $\left(\boldsymbol{\varpi}^{L}, \boldsymbol{\kappa}^{L}\right)$ through the Leapfrog technique. To achieve this, at first run, we utilize the optimized model parameters (OMPs, which refer to the maximum likelihood estimates) of $\boldsymbol{\varpi}$ (say, $\widehat{\boldsymbol{\varpi}}$) and then randomly produce $\boldsymbol{\kappa}$ from $N_{g}(0, 1), g=|\boldsymbol{\kappa}|.$ In the subsequent steps, we adopt the immediately accepted value of $\boldsymbol{\varpi}$. However, for $\boldsymbol{\kappa}$, $g$ independent normal variates $\kappa_{1}, \cdots, \kappa_{g}$ are drawn for each iteration of the HMC-SA. The procedures to sample from the joint posterior distribution using HMC-SA are stated in \textbf{Algorithm \textcolor{red}{1}} .
\newline
\textbf{Algorithm 1}: Posterior sampling from the joint posterior distribution using HMC-SA

	\label{Algth1}
	\small
	\begin{algorithmic}
		\State\textbf{Input:} Specify $\boldsymbol{Q}$, the diagonal matrix with entries $q_{1}, \cdots, q_{g}$, step size $\epsilon$, Leapfrog steps $L$, the simulation size $S$, and then set $\boldsymbol{\varpi}^{0}$, the initial values of $\boldsymbol{\varpi}.$
		
		\State\textbf{Output:} Posterior sample observations  $\boldsymbol{\varpi}^{1}, \boldsymbol{\varpi}^{2}, \cdots, \boldsymbol{\varpi}^{S}.$
		
		\State\textbf{Steps:} 
		
		\State \textbf{1:} For $i=0$ to $S-1$
		\State\textbf{2:} Randomly generate $\boldsymbol{\kappa}$ from $N_{g}(0, \mathbf{Q})$
		\State\textbf{3:} Execute the Leapfrog procedure starting at $ (\boldsymbol{\varpi}^{(\alpha=1)}, \boldsymbol{\kappa}^{(\alpha=1)})$ for $\alpha=1, \cdots, L$ steps with step size $\epsilon$ to iteratively generate the proposed state $ (\boldsymbol{\varpi}^{(L)}, \boldsymbol{\kappa}^{(L)})$
		\State \textbf{4:} Compute the acceptance probability define as
		$$\eta=\min(1,\phi(\boldsymbol{\varpi}, \boldsymbol{\varpi}^{(L)}, \boldsymbol{\kappa}, \boldsymbol{\kappa}^{(L)})).$$
		
		\State \textbf{5:} Generate a uniform random number $\gamma$  from Uniform distribution, i.e., $\gamma\sim \text{Unif}(0,1)$
		\State \textbf{6:} Take
		\begin{equation}	
			\boldsymbol{\varpi}^{i+1}=\left\{
			\begin{aligned}
				&\boldsymbol{\varpi}^{(L)}\quad& \text{if}\; \gamma\le\eta\\
				&\boldsymbol{\varpi}^{i}\quad&\text{elsewhere}.\nonumber
			\end{aligned}
			\right.
		\end{equation}
	\end{algorithmic}
	
		
		
		
	

\section{Formulation of Bayesian Model}\label{S4}
\,\indent In this section, we propose a Bayesian method (BM) for estimating the BFM model parameters and some of the model's reliability characteristics. The study builds the full joint posterior distribution after selecting prior distributions, and uses the HMC-SA to derive posterior samples from the joint posterior distribution.

\subsection{The BFM Likelihood Function}
\,\indent Suppose $\mathbf{x}=(x_{1}, x_{2}, \cdots x_{n})$ is life testing recorded observations from $n$ items and that right censoring is possible. If $\mathcal{X}$ is censored at $x_{i}$, the eventual COF is unknown, so we define the censoring identity function designated by $\varrho_{i}$ as
\begin{equation*}
	\varrho_{i}=\begin{cases}
		1, & \text{if  $x_{i}$ is the minimum FT due to COF 1 or 2}, \\
		0, & \text{if  $x_{i}$ is a censored observation.}
	\end{cases}
\end{equation*}
Under the independent censoring assumption and after a slight simplification, the likelihood function for the BFM density is

\begin{equation}\label{LL}
	\small	\begin{aligned}
		\underline{L}(\mathbf{x}|\boldsymbol{\varpi})&=\prod_{i=1}^{n}{w_{BFM}^{\varrho_{i}}(x_{i})\bar{W}_{BFM}^{1-\varrho_{i}}(x_{i})}\\&=\prod_{i=1}^{n}\left( \frac{\theta\nu x_{i}^{\theta-1}}{ \nu x_{i}^{\theta} +1 }+\frac{\tau\zeta \left( \zeta x_{i}\right) ^{\tau-1}}{e^{-\left( \zeta x_{i}\right) ^{\tau}}}\right)^{\varrho_{i}}\\&\times e^{-\sum_{i=1}^{n}\left\lbrace e^{\left( \zeta x_{i}\right) ^{\tau}}-1+ \ln(\nu x_{i}^{\theta} +1 )  \right\rbrace },
	\end{aligned}
\end{equation}
Although the main estimation tool for analyzing the BFM model in this paper is Bayesian methodology, we additionally propose to use the MLM in the execution of the HMC-SA and for BFM performance evaluation against other competing candidates. The OMPs of the BFM parameters are utilized as the initial values in the execution of HMC-SA. 

\subsection{The prior selection and BFM posterior distribution}
\,\indent Given the fact that all of the BFM parameters have positive support (see \cite{Ranjan2021}), we opt for Gamma prior in the present study. In addition to the parameter positive support, we have also taken into account the flexibility of the Gamma family of distributions with the ability to handle various shapes. Besides that, the positive Bayesian findings in similar recent studies further supported the choice of Gamma priors in this paper. To take some instances, \cite{Al-essa2023} and \cite{Abba2023a} used Gamma distribution as a prior for estimating the parameters of some additive models. The studies demonstrated that the Gamma priors produced accurate Bayesian estimates (BEs) for the parameters of the models. The Gamma distribution is also known as a frequent prior for shape parameters in competing risk models \citep{Zhang2022,Bai2020}. 

Thus, the independent Gamma priors for the BFM parameter, $\nu, \theta, \tau$ and $\zeta$, which are subsequently represented by $\varpi_{h}, h=1, \dots, 4\,\, (\text{i.e., by designating}\,\, \varpi_{1}=\nu,  \varpi_{2}=\theta,  \varpi_{3}=\tau, \varpi_{4}=\zeta)$, are given as
\begin{equation}\label{GP}
	f_{h}(\varpi_{h})\propto
	\varpi_{h}^{a_{h}-1}e^{-b_{h}\varpi_{h}}, \quad a_{h}>0, b_{h}>0, 
\end{equation}
where $a_{h}$ and $b_{h},\,h=1, \cdots, 4$ represent the gamma hyper-parameters and $\frac{b_{h}^{a_{h}}}{\Gamma(a_{h})}$ is the proportionality constant. The expected value and variance of each prior in \eqref{GP} for $h=1, \cdots, 4$ are $a_{h}/b_{h}$ and $a_{h}/b_{h}^{2}$, respectively. With the independent priors being selected, we thus define the joint prior as 
\begin{equation}\label{jgp}
	f(\boldsymbol{\varpi})=f_{1}(\varpi_{1}|\boldsymbol{\mu}_{1})f_{2}(\varpi_{2}|\boldsymbol{\mu}_{2})f_{3}(\varpi_{3}|\boldsymbol{\mu}_{3})f_{4}(\varpi_{4}|\boldsymbol{\mu}_{4}),
\end{equation}
where $\boldsymbol{\mu}_{h}=(a_{h},b_{h})^{\prime}$. Having the function of the likelihood $\underline{L}(\mathbf{x}|\boldsymbol{\varpi})$ in \eqref{LL} and the joint prior distribution $f(\boldsymbol{\varpi})=f(\nu, \theta, \tau, \zeta)$ in \eqref{jgp}, the joint posterior distribution, designated by $\Omega(\boldsymbol{\varpi}|\mathbf{x})$, is derived by incorporating $\underline{L}(\mathbf{x}|\boldsymbol{\varpi})$ and $f(\boldsymbol{\varpi})$ via Bayes' theorem. Thus, we have the joint posterior distribution for the BFM parameters written, up to proportionality, as:
\begin{eqnarray}\label{jp}
	\small	\begin{aligned}	&\Omega(\boldsymbol{\varpi}|\mathbf{x})\propto \varLambda(x|\varrho, \boldsymbol{\varpi})\\&\times\prod_{h=1}^{4}\varpi_{h}^{a_{h}-1}\exp\left[-\sum_{h=1}^{4}b_{h}\varpi_{h}-\sum_{i=1}^{n} R_{BFM}(x_{i},\boldsymbol{\varpi}) \right],
\end{aligned}\end{eqnarray}
where\\ $\small\varLambda(x|\varrho, \boldsymbol{\varpi})=\prod_{i=1}^{n}\left( \frac{\varpi_{1}\varpi_{2} x_{i}^{\varpi_{2}-1}}{ \varpi_{1} t^{\varpi_{2}} +1 }+\frac{\varpi_{3}\varpi_{4} \left( \varpi_{4} x_{i}\right) ^{\varpi_{3}-1}}{e^{-\left( \varpi_{4} x_{i}\right) ^{\varpi_{3}}}}\right)^{\varrho_{i}}, \varrho=(\varrho_{1}, \cdots, \varrho_{n})$, and $ R_{BFM}(x_{i},\boldsymbol{\varpi})$ is given in \eqref{S2E10}.\\
The four marginal posterior densities (MPDs) from \eqref{jp} are obtained as
\begin{eqnarray}\label{fcp1}
	\small	\begin{aligned}
		\Omega&(\varpi_{h}|\mathbf{x})\propto \varLambda(x|\varrho, \boldsymbol{\varpi})\varpi_{h}^{a_{h}-1}\\&\times\exp\left[-b_{h}\varpi_{h}-\sum_{i=1}^{n} \ln\left(\varpi_{1} x_{i}^{\varpi_{2}}+1\right) \right], \text{for}\,\, h=1, 2,\\
		\Omega&(\varpi_{h}|\mathbf{x})\propto \varLambda(x|\varrho, \boldsymbol{\varpi})\varpi_{h}^{a_{h}-1}\\&\times\exp\left[-b_{h}\varpi_{h}-\sum_{i=1}^{n}\left(e^{(\varpi_{4}x_{i})^{\varpi_{3}}} -1\right)  \right], \text{for}\,\, h=3, 4.
		\\
	\end{aligned}
\end{eqnarray}

Adopting the notable error loss function, the square error, we have the estimators of the BFM parameters $\varpi_{h}, h=1, \cdots, 4$, computed by
\begin{eqnarray*}\label{pd}
	\begin{aligned}
		&\widehat{\varpi}_{h}=\mathbb{E}(\varpi_{h}|\mathbf{x})=\int \varpi_{h}\,\Omega(\varpi_{h}|\mathbf{x})\text{d}\varpi_{h}, \quad \text{for}\,\,\, h=1,\dots,4. 
\end{aligned}\end{eqnarray*}
We note that direct determination of the analytical inferential approaches may be infeasible due to the intricate form of the MPDs in \eqref{fcp1}. Consequently, we propose applying the HMC-SA to sample from the $\Omega(\boldsymbol{\varpi}|\mathbf{x})$, as described in \textbf{Algorithm \textcolor{red}{1}}.
Hence, we generate $S$ posterior observations $\left\lbrace \boldsymbol{\varpi}_{i}=(\varpi_{h}^{i}|h=1, \cdots 4\,\, \text{and}\,\,  i=1, \dots, S)\right\rbrace$ from the joint posterior distribution. Moreover, to retain only the stabilized posterior observations, the initial $S_{0}$ sample points are assumed as \textit{burn-in} samples and thus removed. Hence, the rest of $S-S_{0}$ posterior sample points are employed for the posterior analysis. The BEs of the BFM parameters are computed via
\begin{eqnarray*}\label{AE}
	\widehat{\varpi}^{\ast}_{h}\approx\frac{1}{S-S_{0}}\sum_{i=S_{0}+1}^{S}{\varpi_{h}^{i}}, \quad \text{for}\,\,\, h=1,\cdots,4. \end{eqnarray*}

In Bayesian analysis, it is empirically evidenced that running several parallel chains, for example, $\iota_{0}=3, 4,$ or $5$, broadly facilitates the sampler convergence compared to employing a single chain. Therefore, for $\iota_{0}$ parallel chains, the BEs are calculated from
\begin{eqnarray*}\label{AES}
	\widehat{\varpi}^{\ast}_{h}\approx\frac{1}{\bar{r}(S-S_{0})}\sum_{\bar{\iota}_{0}=1}^{\iota_{0}}\sum_{i=S_{0}+1}^{S}{\varpi^{\bar{\iota}_{0},i}_{h}}, \quad \text{for}\,\,\, h=1,\cdots,4. 
\end{eqnarray*}
The HMC-SA posterior sample convergence is examined for each parallel chain through the \textit{Potential Scale Reduction Factor} (PSRF) denoted by R-hat \citep{Gelman2015}. It is determined as the weighted mean of the variances between chains and within chains. A posterior sample from the HMC-SA converges if $Rhat<1.1$, and thus divergence exists if otherwise.

\subsection{Implementing the HMC-SA for BFM Posterior Sampling}
\,\indent The HMC-SA implementation requires the first partial derivative of the logarithmic joint posterior distribution (i.e., $\ln\Omega(\boldsymbol{\varpi}|\mathbf{x})$) with respect to the $\boldsymbol{\varpi}_{h}, h=1, \cdots, 4$, as obviously shown by Equation \eqref{S3E5}. This is equal to adding the derivative of  the natural logarithm of the likelihood function and joint prior, provided as 
\begin{equation*}\label{S4E1}
	\frac{\partial \ln\left( \Omega(\boldsymbol{\varpi}|\mathbf{x})\right) }{\partial\boldsymbol{\varpi}_{h}}=	\frac{\partial  \ln\left(\underline{L}(\mathbf{x}|\boldsymbol{\varpi})\right) }{\partial\boldsymbol{\varpi}_{h}}+	\frac{\partial  \ln\left(f(\boldsymbol{\varpi})\right) }{\partial\boldsymbol{\varpi}_{h}}.
\end{equation*}
Thus, for the BFM model, we have the derivative of $\ln\Omega(\boldsymbol{\varpi}|\mathbf{x})$) with respective to $\boldsymbol{\varpi}_{h}$, given as
\begin{equation}\label{S4E2}
	\small	\begin{aligned}
		&\frac{\partial \ln\Omega(\boldsymbol{\varpi}|\mathbf{x}) }{\partial\boldsymbol{\varpi}_{h}}=	\frac{\partial   }{\partial\boldsymbol{\varpi}_{h}}\sum_{i=1}^{n}\varrho_{i}\ln\left( \tilde{\varLambda}(x| \boldsymbol{\varpi})\right) -\sum_{i=1}^{n}\frac{\partial   }{\partial\boldsymbol{\varpi}_{h}}\left(  e^{\left( \varpi_{4} x_{i}\right) ^{\varpi_{3}}}-1+ \ln(\varpi_{1} x_{i}^{\varpi_{2}} +1 )  \right)  \\&+	\frac{\partial   }{\partial\boldsymbol{\varpi}_{h}}\ln\left(\frac{b_{1}^{a_{1}}}{\Gamma(a_{1})}\frac{b_{2}^{a_{2}}}{\Gamma(a_{2})}\frac{b_{3}^{a_{3}}}{\Gamma(a_{3})}\frac{b_{4}^{a_{4}}}{\Gamma(a_{4})}\right)
		+	\frac{\partial   }{\partial\boldsymbol{\varpi}_{h}}\ln\left( 	\varpi_{1}^{a_{1}-1}\varpi_{2}^{a_{2}-1}\varpi_{3}^{a_{3}-1}\varpi_{4}^{a_{4}-1} \right)
		\\&-	\frac{\partial   }{\partial\boldsymbol{\varpi}_{h}}\left( 	b_{1}\varpi_{1}+b_{2}\varpi_{2}+b_{3}\varpi_{3}+b_{4}\varpi_{4} \right),
	\end{aligned}
\end{equation}
where $\tilde{\varLambda}(x| \boldsymbol{\varpi})= \frac{\varpi_{1}\varpi_{2} x_{i}^{\varpi_{2}-1}}{ \varpi_{1} t^{\varpi_{2}} +1 }+\frac{\varpi_{3}\varpi_{4} \left( \varpi_{4} x_{i}\right) ^{\varpi_{3}-1}}{e^{-\left( \varpi_{4} x_{i}\right) ^{\varpi_{3}}}}$. The first partial derivative of \eqref{S4E2} with respect to the $\boldsymbol{\varpi}_{h}, h=1, \cdots, 4,$ are expressed as
\begin{equation}\label{S4E3}
	\small	\begin{aligned}
		&\frac{\partial \ln\Omega(\boldsymbol{\varpi}|\mathbf{x})}{\partial\boldsymbol{\varpi}_{1}}=\sum_{i=1}^{n}\frac{-\varrho_{i}\varpi_{2}x_{i}^{\varpi_{2}-1}}{\tilde{\varLambda}(x| \boldsymbol{\varpi})\left( \varpi_{1}x_{i}^{\varpi_{2}}+1\right) ^{2}}-\sum_{i=1}^{n}\frac{x_{i}^{\varpi_{2}}}{\varpi_{1}x_{i}^{\varpi_{2}}+1}+\frac{a_{1}-1}{\varpi_{1}}-b_{1},\\
		&\frac{\partial \ln\Omega(\boldsymbol{\varpi}|\mathbf{x})}{\partial\boldsymbol{\varpi}_{2}}=\sum_{i=1}^{n}\frac{\varrho_{i}\varpi_{1}\left\lbrace \varpi_{2}\ln(x_{i})+1\right\rbrace x_{i}^{\varpi_{2}-1}}{\tilde{\varLambda}(x| \boldsymbol{\varpi})\left( \varpi_{1}x_{i}^{\varpi_{2}}+1\right) ^{2}}-\sum_{i=1}^{n}\frac{ \varpi_{1}x_{i}^{\varpi_{2}}\ln(x_{i})}{\varpi_{1}x_{i}^{\varpi_{2}}+1}+\frac{a_{2}-1}{\varpi_{2}}-b_{2},\\
		&\frac{\partial \ln\Omega(\boldsymbol{\varpi}|\mathbf{x})}{\partial\boldsymbol{\varpi}_{3}}=\sum_{i=1}^{n}\frac{\varrho_{i}\varpi_{4}\left\lbrace\varpi_{3}\left[(\varpi_{4}x_{i})^{\varpi_{3}}+1 \right]\ln(\varpi_{4}x_{i})+1  \right\rbrace }{\tilde{\varLambda}(x| \boldsymbol{\varpi})(\varpi_{4}x_{i})^{1-\varpi_{3}}e^{-(\varpi_{4}x_{i})^{\varpi_{3}}}}\\&-\sum_{i=1}^{n}{\frac{(\varpi_{4}x_{i})^{\varpi_{3}}\ln(\varpi_{4}x_{i})}{e^{-(\varpi_{4}x_{i})^{\varpi_{3}}}}}+\frac{a_{3}-1}{\varpi_{3}}-b_{3},\\
		&\frac{\partial \ln\Omega(\boldsymbol{\varpi}|\mathbf{x})}{\partial\boldsymbol{\varpi}_{4}}=\sum_{i=1}^{n}\frac{\varrho_{i}\varpi_{3}^{2}(\varpi_{4}x_{i})^{\varpi_{3}-1}\left\lbrace(\varpi_{4}x_{i})^{\varpi_{3}}+1  \right\rbrace }{\tilde{\varLambda}(x| \boldsymbol{\varpi})e^{-(\varpi_{4}x_{i})^{\varpi_{3}}}}-\sum_{i=1}^{n}{\frac{\varpi_{3}x_{i}(\varpi_{4}x_{i})^{\varpi_{3}-1}}{e^{-(\varpi_{4}x_{i})^{\varpi_{3}}}}}+\frac{a_{3}-1}{\varpi_{3}}-b_{3}.
	\end{aligned}
\end{equation}
Noting that the BFM parameters are limited to positive support of the real space and the HMC-SA requires the complete real space, we employ logarithmic transformation to redefine the parameter support as follows.
\begin{equation}\label{S4E4}
	\begin{aligned}
		\underline{\varpi}_{h}=\ln(\varpi_{h})\,\, \text{for}\,\, h=1, \cdots, 4.
	\end{aligned}
\end{equation}
Employing the transformed parameters \eqref{S4E4}, we redefine \eqref{S4E3} to have the following relationships.
\begin{equation}\label{S4E5}
	\small	\begin{aligned}
		&\frac{\partial \ln\left( \Omega(\boldsymbol{\underline{\varpi}}|\mathbf{x})\right) }{\partial\boldsymbol{\underline{\varpi}}_{1}}=	\left( \frac{\partial  \ln\left(\underline{L}(\mathbf{x}|\boldsymbol{\underline{\varpi}})\right) }{\partial\boldsymbol{\underline{\varpi}}_{1}}+	\frac{\partial  \ln\left(f(\boldsymbol{\underline{\varpi}})\right) }{\partial\boldsymbol{\underline{\varpi}}_{1}}\right) \frac{\partial  \varpi_{1} }{\partial\boldsymbol{\underline{\varpi}}_{1}},\\
		&\text{for}\,\, h=1, \cdots, 4.
\end{aligned} \end{equation}
Equation \eqref{S3E5} can, therefore, be applied to determine the proposal points for the pre-specified iterating steps once the derivatives in \eqref{S4E5} are realized. For each iterating step, the acceptance probability ($\eta$) in \eqref{S3E6} can then be employed to accept or reject the proposed point, as discussed in Section \ref{S3}. Hence, as described in \textbf{Algorithm \textcolor{red}{1}}, the HMC-SA can be executed once every detail is offered.

It can be seen from \eqref{S4E2} to \eqref{S4E5} that deriving and handling the first partial derivatives $\frac{\partial \ln\left( \Omega(\boldsymbol{\underline{\varpi}}|\mathbf{x})\right) }{\partial\boldsymbol{\underline{\varpi}}_{h}}$ could be problematic depending on the intricate form of the joint posterior distribution under study. Moreover, besides the selection of $\boldsymbol{Q}$ - the momentum distribution in the algorithm, the HMC-SA efficiency is also not independent of the choice of tuning parameters per iteration, that is, the scaling factor $\epsilon$ and the Leapfrog steps $L$, which required manual adjustments. Consequently, to ease the computational intricacy of the HMC-SA, \cite{Hoffman2014} introduced the No-U-Turn Sampler (NUTS) - an alternate version of the HMC-SA. The NUTS reduces the HMC-SA burden by automatically updating the tuning parameters during the warm-up period, and are then held fixed throughout the subsequent iterations. Interestingly, we presently have an open-source software called \verb|Stan| \citep{Team2018}, which incorporates the NUTS developments. Thus, in this article, we use the \verb|RStan| - an \verb|R| package interface to the \verb|Stan| for posterior sampling from the joint posterior distribution.

The \verb|Stan| uses the reverse-mode automatic differentiation \citep{Griewank2008} to compute the derivative of the log-posterior with respect to the model's parameters. In this approach, the Stan's compiler utilizes the function provided by the Stan program and computes analytically its derivative through an efficient method. This procedure is believed to be faster than numerical differentiation, particularly in models with many parameters. Thus, a solution to the gradient or derivatives of the log-posterior complexity.

\section{Illustrations with Competing Risk Data}\label{S5}
\,\indent Illustrating the proposed BFM model's good performance in real applications remains a critical aspect of the present study. To accomplish this, we select two CR datasets with NMFR features, comprising the 58 electrodes' failure and survival times (EFST) \citep{Doganaksoy2002} and the 33 electrical appliances' failure and survival times (AFST) \citep{Lawless2003}. Both datasets are identified with two or more COFs. The TTT-transform graphs in \figurename{s \ref{figTTT}}(I) and \ref{figTTT}(II) verified the FR features of the FT distributions for the cause-specific and combined FTs for the two datasets, as discussed in Section \ref{S1}. These datasets are observed to have similar FR behaviors with the BFM model (i.e., BFR for EFST and RCFR for AFST) and thus suggest the appropriateness of the BFM for analyzing the datasets. For each illustration (see Sections \ref{S6.1} and \ref{S6.2}), we employ the HMC-SA to determine the Bayesian estimates of the BFM parameters. We run $S=$2000 iterations with four parallel chains for the HMC-SA. Of the $S$ iterations, we retain the last $S-S_{0}=1000$ posterior observations for posterior analysis.  The performance evaluation of the proposed BFM against several recent CR models and other competing candidates (see \tablename{ \ref{model_tab1}}) is subsequently performed under the MLM. In this case, four parametric metrics (PM) and three non-parametric metrics (NPM) are involved, comprising the negative loglikelihood value (PM$_{-\text{log}\underline{L}}$), Akaike information criterion (PM$_{AIC}$), Bayesian information criterion (PM$_{BIC}$), Bridge criterion (PM$_{BC}$), Kolmogorov-Smirnov (NPM$_{KS}$), Anderson-Darling (NPM$_{AD}$), and Cramer-von Misses (NPM$_{CVM}$). We also provide reliability function estimates and graphs to support the performance assessments. 
\begin{table}[h]
	\caption{\small FR Behavior of BFM and five other competing candidates.}
	\label{model_tab1}
	\centering
	\begin{adjustbox}{width=4.7in}
		\begin{tabular}{lc}\toprule
			\large
			Model & FRF ($r(x)$) and its behaviors\\
			\midrule
			\textbf{Proposed BFM} &$\frac{\theta\nu x^{\theta-1}}{ \nu x^{\theta} +1 }+\frac{\tau\zeta \left( \zeta x\right) ^{\tau-1}}{e^{-\left( \zeta x\right) ^{\tau}}}$\\
			& IBBFR, RCFR, BFR\\
			Additive Perks distribution (APD)\citep{Mendez-Gonzalez2022a}&$\frac{\alpha\lambda e^{\lambda x}}{1+\alpha e^{\lambda x}}+\frac{\beta\theta e^{\theta x}}{1+\beta e^{\theta x}}$\\
			&BFR\\
			Flexible additive Chen-Gompertz (FACG)\citep{Abba2022}	&$\alpha\gamma x^{\gamma-1}e^{x^{\gamma}}+\lambda e^{\lambda x-\theta}$\\	
			&BFR\\
			Flexible additive exponential power-Gompertz (FAEPG)\citep{Abba2023a}	&	$\tau\upsilon  \left(\upsilon x \right)^{\tau-1}e^{\left(\upsilon x \right) ^{\tau} }+\alpha e^{\alpha x-\theta}$\\
			&BFR\\
			Exponentiated additive Weibull distribution (EAddW)\citep{Ahmad2020}&$\frac{\theta e^{-\alpha x^{\beta}-\gamma x^{\lambda}}\left(\alpha\beta x^{\beta-1}+\gamma\lambda x^{\lambda-1} \right) }{ \left(1-e^{-\alpha x^{\beta}-\gamma x^{\lambda}}\right)^{-\theta+1}\left[ 1-\left(1-e^{-\alpha x^{\beta}-\gamma x^{\lambda}}\right)^{\theta}\right] }$\\
			& IBFR  and BFR\\
			Generalized extended exponential-Weibull (GExtEW)\citep{Shakhatreh2020}&$\frac{c\alpha(\gamma\beta x^{\gamma-1}+\lambda)e^{-(\beta(x^\gamma) +\lambda  x)^c} (1-e^{-(\beta(x^\gamma) +\lambda x)^c})^{\alpha-1}}{(\beta(x^\gamma) +\lambda x)^{1-c}\left( 1-(1-e^{-(\beta(x^\gamma) +\lambda x)^c})^\alpha\right) }$\\
			&IBFR, BIBFR, IBBFR, BFR\\
			\bottomrule	
		\end{tabular}
	\end{adjustbox}
\end{table}
\subsection{Illustration I: Electrodes Failure and Survival Times}\label{S6.1}
\,\indent This data set portrays the EFST (in hours) of electrodes \citep{Doganaksoy2002}, which comprises 58 electrodes subjected to a high-voltage endurance life test. The electrodes' failures were observed to have been caused by insulation defects associated with a processing problem ($C_{1}$) or degradation of the organic material ($C_{2}$). Of the recorded failures, 18 of the EFSTs were caused by $C_{1}$, 27 of the failures happened due to $C_{2}$, and the remaining 13 electrodes were still working after the test ended and were assumed censored.

The data have BFR, IFR, and BFR curves, for accordingly, $C_{1}$, $C_{2}$ and combined data, as portrayed in \figurename{ \ref{figTTT}}. 
After the Bayesian modeling of the data using the BFM model, we subsequently evaluate the model's performance against the models in \tablename{ \ref{model_tab1}}. We note that the APD, FACG, and FAEPG are CR models with BFR characteristics, while the EAddW and GExtEW are extensions of other CR models. 

\subsubsection{Posterior Estimates and Plots}\label{S6_1_1}
\,\indent For the chosen priors' hyper-parameters, $a_{h}$ and $b_{h}$ for $h=1,\cdots,4$, we assigned values that approximately correspond to the averages of the BFM parameters' OMPs. With these hyper-parameter values' selection, we executed the HMC-SA as detailed in Section \ref{S5}. In order to enable better sampler convergence, $\iota_{0}=4$ parallel chains, each having $S=$2000 iterations were run. We retain the last $S-S_{0}=1000$ posterior observations for posterior diagnostic graphs and summaries for each parallel chain.

\begin{figure}[h!]
	\centering	
	\includegraphics[width=3.3in, height=3.25in, keepaspectratio=false]{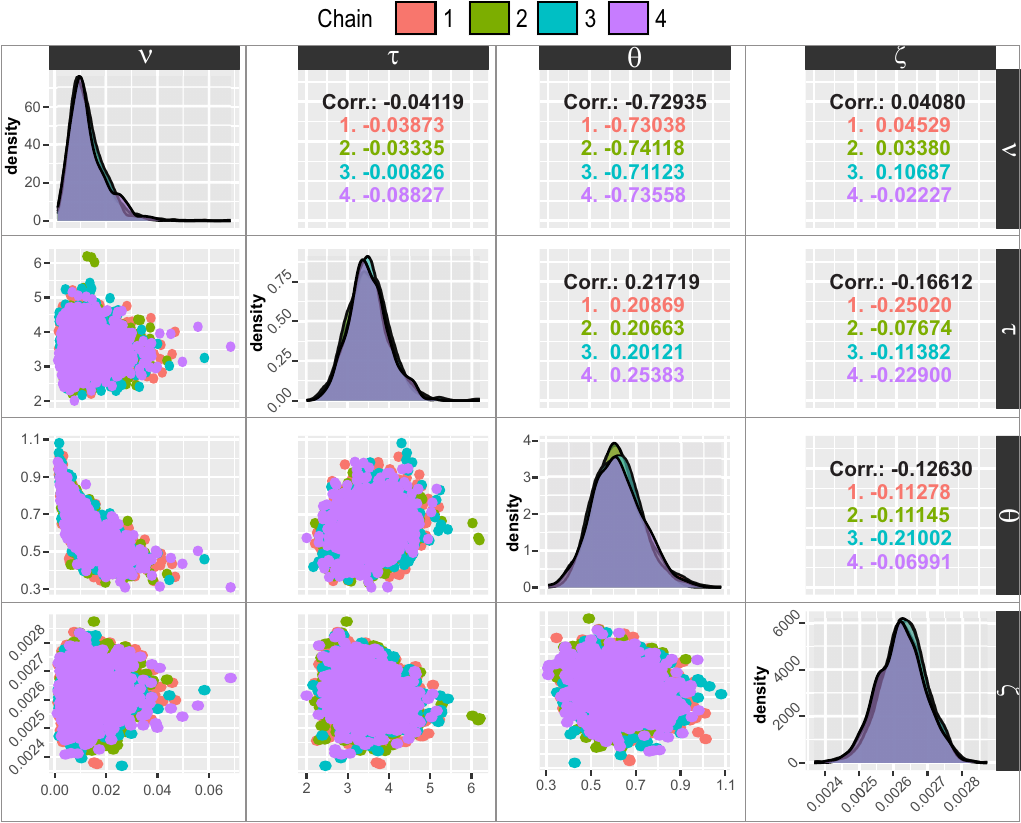}
	\centering
	\caption{\small Scatter graph depicting (\textit{up-diagonal}) posterior correlation estimates, (\textit{diagonal}) posterior density estimates, and (\textit{below-diagonal}) scatter diagram between the bi-variate of the BFM parameters posterior samples; EFST data.}
	\label{EFST_fig1}
\end{figure}

We provide the posterior diagnostic graph in \figurename{ \ref{EFST_fig1}}. From below and up the diagonal, the figure depicts the $\iota_{0}=4$ parallel chains scatter graphs between the bi-variate posterior sample pairs and bi-variate posterior correlation estimates, respectively. It can be observed that the parameters ($\nu$ and $\theta$) from the Dhillon component in the BFM model are highly correlated, which can also be seen from the related scatter diagram. These parameters have, however, recorded weak correlation estimates with parameters ($\zeta$ and $\tau$) from the exponential power component. Likewise, the estimated correlation between the exponential power component parameters $\zeta$ and $\tau$ is -0.16612 on average, translating to less dependence between the two. To learn more about how well HMC-SA works, we see that the kernel estimates show single-modal and bell-shaped posterior densities for most of the BFM parameters, as seen in the diagonal of \figurename{ \ref{EFST_fig1}}. Additionally, the $R-hat$ estimates of 0.9999, 1.0014, 1.0008, and 1.0003 for accordingly, $\nu$, $\theta$, $\tau$, and $\zeta$, have obviously ascertained the MCMC sample convergences (\tablename{ \ref{EFST_tab1}}).

We numerically estimate the BFM parameters and the highest posterior density intervals (HPDIs) by applying the generated posterior samples. Under the MLM, we computed the OMPs and asymptotic confidence intervals (ACIs) of the BFM parameters. The BEs and OMPs are reported in \tablename{ \ref{EFST_tab1}}. It can be seen that the BFM parameter estimates from the two estimated techniques are very similar to each other, with minimal deviation in $\tau$. The BM has given more minimal  St-Devs of 0.000145, 0.0025, 0.0099, and 0.000001 compared to the MLM with St-Devs of 0.0111, 0.1818, 0.7377, and 0.000097, for the $\nu$, $\theta$, $\tau$ and $\zeta$, respectively. These minimal St-Devs have led to narrowed-width  HPDIs for the BEs, which is better than the ACIs. Next, we compare the proposed BM and MLM based on the three NPM values, as presented in \tablename{ \ref{EFST_tab1}} for the two approaches. It can be noticed that BM has better strength in estimating the BFM parameters based on all three comparison metrics. The metrics gave the highest \textit{P}-values under the BM. The MTTF estimates under the BM and MLM are also given in \tablename{ \ref{EFST_tab1}}.

\begin{table}[h!]
	\caption{\small The BEs, OMPs, St-Devs, 95\% HPDIs, and ACIs of the BFM parameters for modeling EFST data. The three NPMs are provided to assess the two approaches.} 
	\label{EFST_tab1} 
	\centering
	\begin{adjustbox}{width=4.5in}
		\begin{tabular}{lcccc} \toprule 
			
			Parameter	& $	\widehat{\nu}$	&$	\widehat{\theta}$ &$	\widehat{\tau}$&$	\widehat{\zeta}$\\	\midrule
			
			\textbf{BE (St-Dev)}	&0.0128 (1.45$e^{-4}$)&0.6172 (0.0025)&3.4994 (0.0099)&0.0026 (1.00$e^{-6}$)\\	
			95\% HPD Interval&[0.0021, 0.0260]&[0.4087, 0.8349]&[2.6097, 4.5390]&[0.0025, 0.0028]\\
			R-hat&0.9999&1.0014&1.0008&1.0003\\	\cline{2-5}
			NPM$_{\text{KS}}(P_{\text{KS}})$&0.1443 (0.1784)&&&\\
			NPM$_{\text{AD}}(P_{\text{AD}})$&0.1597 (0.9464)&&$\widehat{\text{MTTF}}_{\text{BM}}\rightarrow$&241.89\\
			NPM$_{\text{CvM}}(P_{\text{CVM}})$&0.0232 (0.9307)&&&\\
			\cline{2-5}	
			\textbf{	OMP (St-Dev)} &0.0127 (0.0111)&0.6124 (0.1818)&3.5770 (0.7377)&0.0026 (9.72$e^{-5}$)\\		
			95\% AC Interval&[0.0009,  0.0345]&[0.2561, 0.9687]&[2.1307, 5.0224]&[0.0024, 0.0028]\\	\cline{2-5}
			NPM$_{\text{KS}}(P_{\text{KS}})$&0.1454 (0.1723)&&&\\
			NPM$_{\text{AD}}(P_{\text{AD}})$&0.1993 (0.8798)&&$\widehat{\text{MTTF}}_{\text{MLM}}\rightarrow$&243.88\\
			NPM$_{\text{CvM}}(P_{\text{CVM}})$&0.0289 (0.8583)&&&\\ \bottomrule
		\end{tabular}
	\end{adjustbox}
\end{table}

\subsubsection{BFM Performance Against Some Recent CR Models}
\,\indent In this segment, we evaluate the robustness of the BFM model using MLM against five other CR models and its related extension for the analysis of EFST data. The OMPs of the models are provided in \tablename{ \ref{EFST_tab2}}. For assessing the BFM fitting performance related to the five competing candidates, the seven metrics are reported in \tablename{ \ref{EFST_tab2}}. We additionally determined the ranks of the metrics, denoted by $\alpha_{R}$, to support the models' assessment.

From the metrics, the trained BFM model is selected as a better model for describing the EFST data. The chosen BFM corresponds to the minimum values of six out of the seven metrics as boldly marked in \tablename{ \ref{EFST_tab2}}. That is, apart from the trained APD model, which took the minimum NPM$_{\text{AD}}$ value, the BFM indicates a better fit to the EFST data based on all four PMs and two NPMs. The average rank-wise assessment of these metrics portrays the BFM in the leading position as shown in the 10$-th$ column of \tablename{ \ref{EFST_tab2}}. Therefore, the proposed BFM establishes its potential over the APD, FACG, FAEPG, EAddW and GExtEW models. While the FAEPG is numerically seen as the worst-performing model, the APD has comforted itself as the second-best-performing model.  

\tablename{ \ref{EFST_tab3}} shows the estimated risks of failure ($\widehat{F}_{C_{1}}$ and $\widehat{F}_{C_{2}}$) caused by the insulation defects ($C_{1}$) and degradation of the organic material ($C_{2}$) for the BFM and three other trained CR models, including the APD, FACG and FAEPG models. The table also reported the empirical probabilities of failure, which are 0.60 and 0.40, respectively, due to $C_{1}$ and $C_{2}$. Results from the trained BFM under BM (i.e., BFM-BM) have attributed 69.8\% to $C_{1}$ and  30.2\%  to $C_{2}$, which are closer to the empirical values of 60\% and 40\% than the estimated values computed from other competing models.


\begin{table}[h!]
	\caption{\small The OMPs with St-Devs within parenthesis and performance metrics of BFM and other competing candidates for EFST data fitting.} 
	\label{EFST_tab2} 
	\centering
	\begin{adjustbox}{width=4.5in}
		\begin{tabular}{cccccccccc} \toprule 
			\large
			\underline{	Model}	&\underline{ML estimate(St-Dev)} &\multicolumn{7}{c}{\underline{Statistics}}&	\\  
			$\downarrow$		&$\downarrow$&PM$_{-\text{log}\underline{L}}$ &PM$_{\text{AIC}}$&PM$_{\text{BIC}}$&PM$_{\text{BC}}$&NPM$_{\text{KS}}$&NPM$_{\text{AD}}$&NPM$_{\text{CVM}}$&$\alpha_{R}$	\\ 		
			&& &&&&$(P_{\text{KS}})$&$(P_{\text{AD}})$&$(P_{\text{CVM}})$&\\	\cline{3-9}
			BFM	& $\widehat{\nu}=0.013(0.011),   $&\textbf{274.79}$\color{red}{^{1}}$&\textbf{557.59}$\color{red}{^{1}}$&\textbf{565.83}$\color{red}{^{1}}$&\textbf{580.81}$\color{red}{^{1}}$&\textbf{0.145}$\color{red}{^{1}}$&0.199$\color{red}{^{2}}$&\textbf{0.029}$\color{red}{^{1}}$&$\color{red}{\approx1}$	\\ 
			&$\widehat{\tau}=3.577(0.738),  $&&&&&(0.172)&(0.880)&(0.858)&\\ 
			&$ \widehat{\theta}=0.612(0.182),$&&&&&&&&\\ 
			&$ \widehat{\zeta}=0.003(9.7e^{-5})$&&&&&&&&\\ 
			APD	&$\widehat{\alpha}=4.5e^{-4}(1.4e^{-3}),  $ &275.78$\color{red}{^{2}}$&559.57$\color{red}{^{2}}$&567.81$\color{red}{^{2}}$&582.78$\color{red}{^{2}}$&0.175$\color{red}{^{4}}$&\textbf{0.184}$\color{red}{^{1}}$&0.033$\color{red}{^{2}}$	&$\color{red}{\approx2}$\\ 
			&$\widehat{\beta}=0.395(0.387), $ 
			&&&&&(0.057)&(0.904)&(0.803)&\\
			&$ \widehat{\lambda}=0.025(0.008), $&&&&&&&&\\ 
			&$\widehat{\theta}=0.012(0.0135)  $&&&&&&&&\\ 
			FACG	&$\widehat{\gamma}=0.323(0.016),  $ &282.21$\color{red}{^{5}}$&572.41$\color{red}{^{5}}$&580.65$\color{red}{^{5}}$&595.628$\color{red}{^{5}}$&0.165$\color{red}{^{3}}$&0.933$\color{red}{^{5}}$&0.159$\color{red}{^{5}}$&$\color{red}{\approx5}$	\\ 
			&$ \widehat{\alpha}=0.002(0.001),$ &&&&&(0.086)&(0.017)&(0.018)&	\\ 
			&$\widehat{\lambda}=0.224(0.008),   $&&&&&&&&\\ 
			&$\widehat{\theta}=99.93(3.593) $&&&&&&&&\\ 		
			FAEPG	&$\widehat{\alpha}=1.244(0.166),  $ &282.44$\color{red}{^{6}}$&572.88$\color{red}{^{6}}$&581.13$\color{red}{^{6}}$&596.10$\color{red}{^{6}}$&0.190$\color{red}{^{5}}$&1.335$\color{red}{^{6}}$&0.229$\color{red}{^{6}}$	&$\color{red}{\approx6}$\\ 
			&$\widehat{\gamma}=0.027(2.2e^{-4}), $ 
			&&&&&(0.030)&(0.002)&(0.002)&\\
			&$\widehat{\lambda}=1.583(0.013),  $&&&&&&&&\\ 
			&$\widehat{\theta}=705.9(0.121)  $&&&&&&&&\\ 
			EaddW&$\widehat{\alpha}=0.831(4.0e^{-3}),  $ &278.21$\color{red}{^{4}}$&566.42$\color{red}{^{4}}$&576.72$\color{red}{^{4}}$&590.63$\color{red}{^{4}}$&0.150$\color{red}{^{2}}$&0.664$\color{red}{^{4}}$&0.112$\color{red}{^{4}}$	&$\color{red}{\approx4}$\\ 
			&$ \widehat{\beta}=0.109(0.017), $ 
			&&&&&(0.145)&(0.079)&(0.075)&\\
			&$\widehat{\gamma}=1.4e^{-6}(2.2e^{-7}), $&&&&&&&&\\
			&$\widehat{\lambda}=2.403(1.2e^{-3}), $&&&&&&&&\\ 
			&$\widehat{\theta}=7.128(8.9e^{-4}) $&&&&&&&&\\ 
			
			GExtEW&$\widehat{\alpha}=0.206(0.029),  $ &276.56$\color{red}{^{3}}$&563.13$\color{red}{^{3}}$&573.43$\color{red}{^{3}}$&587.34$\color{red}{^{3}}$&0.206$\color{red}{^{6}}$&0.337$\color{red}{^{3}}$&0.055$\color{red}{^{3}}$&$\color{red}{\approx3}$	\\ 
			&$ \widehat{\beta}=0.441(7.3e^{-3}), $ 
			&&&&&(0.015)&(0.493)&(0.432)&\\
			&$\widehat{\gamma}=0.080(-), ,$&&&&&&&&\\
			&$\widehat{\lambda}=7.4e^{-4}(5.5e^{-5}) $&&&&&&&&\\ 
			&$\widehat{c}=25.46(0.052) $&&&&&&&&\\ 
			\bottomrule
		\end{tabular}
	\end{adjustbox}
	
\end{table}	

\begin{table}[h]
	\caption{\small Empirical and estimated probabilities of failures due to $C_{1}$ and $C_{2}$ for the BFM and other competing models; EFST data.} 
	\label{EFST_tab3} 
	\centering
	\begin{adjustbox}{width=3.3in}
		\begin{tabular}{lcccccc} \toprule 
			&Empirical&BFM-BM &BFM-MLM&APD&FACG&FAEPG\\
			\cline{2-7}
			$	\widehat{F}_{C_{1}}$	&0.600&0.698&0.705&0.728&0.927&0.926	\\
			$	\widehat{F}_{C_{2}}$	 &0.400&0.302&0.295&0.272&0.073&0.074	\\
			\bottomrule
		\end{tabular}
	\end{adjustbox}
\end{table}

\subsubsection{Model Compatibility}
\,\indent
To ensure that the proposed BFM model is compatible with the EFST data, we employ a simple yet informal strategy to investigate the observed and predicted samples based on the graphs of the BFM reliability functions (for more details, see \cite{Wang2022}). 

\begin{figure}[h!]
	\centering	
	\includegraphics[width=4.5in, keepaspectratio=false]{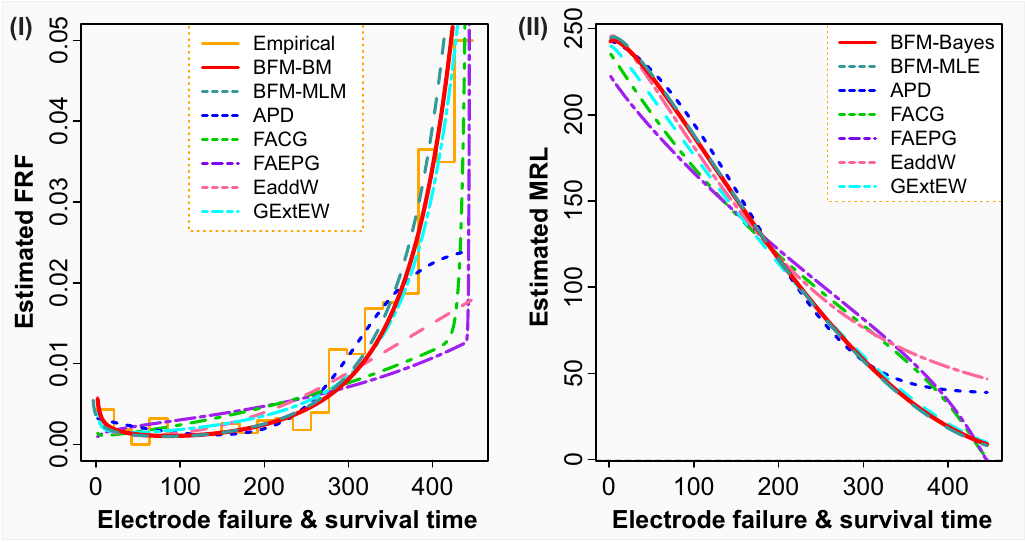}
	\centering
	\caption{\small (I) Empirical and estimated FRFs and (II) estimated MRLFs of BFM and other candidates, portraying the reciprocal relationship between the two reliability features on modeling EFST data.}
	\label{EFST_fig2}
\end{figure}

\figurename{s \ref{EFST_fig2}(I)} and \ref{EFST_fig2}(II) visualize the FRFs $r(x)$ and MRLFs $\mu(\tilde{x})$, respectively, of the trained models on EFST data to evaluate the compatibility of the established BFM against the five competing candidates.  \figurename{s \ref{EFST_fig2}(I)} also portrays the empirical FR of the data. The estimated FR curves of the proposed BFM model under the BM and MLM have fundamentally the same features and nicely mimic the empirical FR of the EFST data. During the useful-life period of the electrode ($x\approx20$ to $250$), the BFM predicts a lower FR than most of its competing candidates. It is noted that within this period, the trained APD predicts a relatively lower FR than the BFM immediately after $x\approx140$ to $240$, which felt below the BFM's coverage of the useful-life period. The graph also shows how the BFM correctly predicts IFR at the wear-out period better than other competing candidates. Although the decreasing of the empirical and predicted FRs in \figurename{ \ref{EFST_fig2}(I)} is very short, the BFM has still managed to predict the reciprocal relationship between its trained FRF and MRLF as shown in  \figurename{s \ref{EFST_fig2}(I)} and \ref{EFST_fig2}(II). While the other trained models have failed to identify this relationship in \figurename{s \ref{EFST_fig2}(II)}, the BFM and EAddW models have predicted a very slight increasing pattern in the early period, which is immediately followed by a decreasing shape.

\begin{figure}[h!]
	\centering	
	\includegraphics[width=4.5in, height=3.2in, keepaspectratio=false]{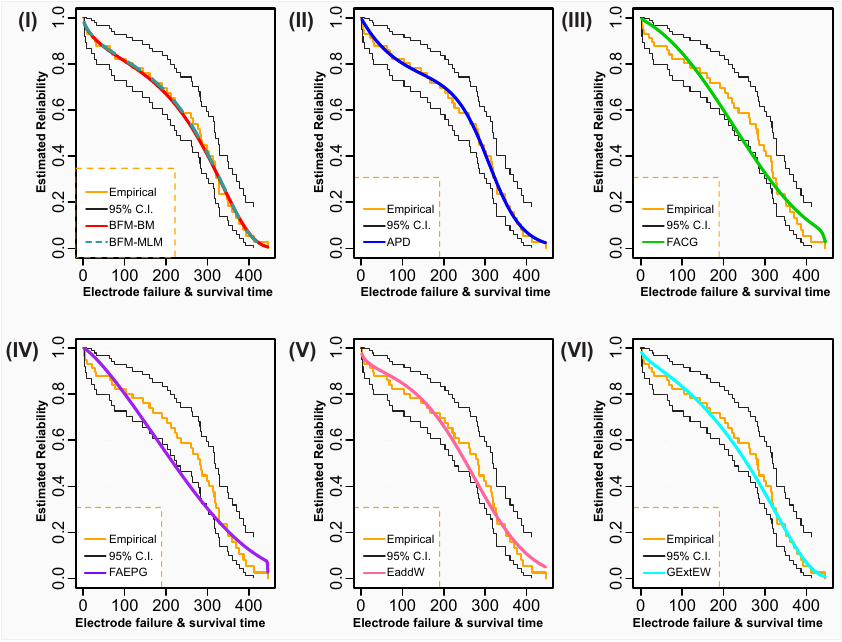}
	\centering
	\caption{\small Empirical and reliability estimates of BFM and five competing models; EFST data.}
	\label{EFST_fig3}
\end{figure}

Besides visualizing the FRFs and MRLFs of the trained models, we provide the empirical and fitted reliability functions of these models in  \figurename{s \ref{EFST_fig3}(I)}--\ref{EFST_fig3}(VI). It can be noticed that the reliability curves of the BFM model (as shown in \figurename{ \ref{EFST_fig3}(I)}) follow the empirical distribution much more closely than the other competing models. Arguably, the excellent performance of the BFM, both numerically and visually, may be related to the flexibility contributions given the Dhillon and exponential power components of the model.


\begin{figure}[h!]
	\centering	
	\includegraphics[width=3.5in, keepaspectratio=false]{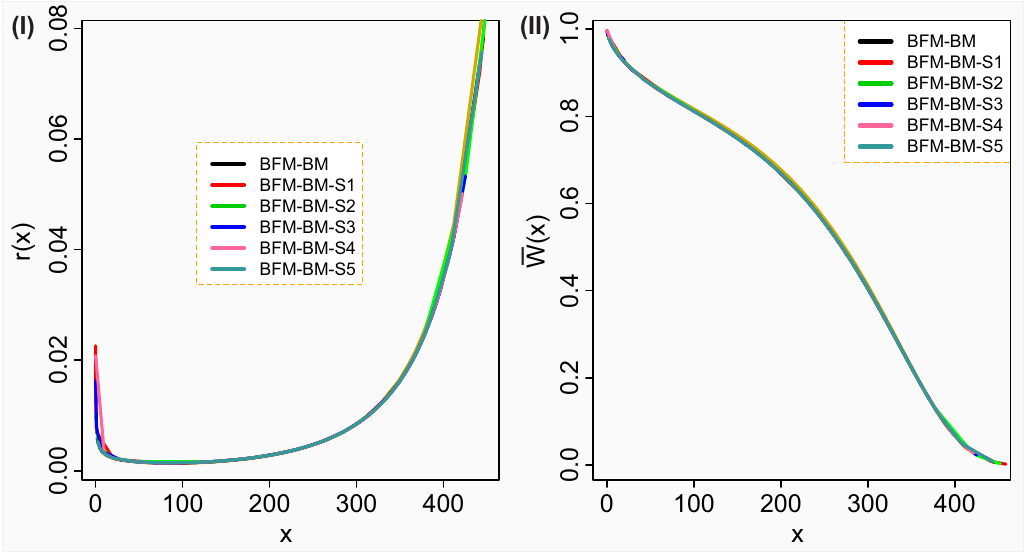}
	\centering
	\caption{\small (I) Empirical and estimated FRFs and (II) estimated MRLFs of BFM and other candidates, portraying the reciprocal relationship between the two reliability features on modeling EFST data.}
	\label{EFST_fig4}
\end{figure}

To further explore the compatibility of the proposed BFM for the Bayesian modeling of the EFST data, five different random samples of equal size with the observed data were drawn from the trained BFM model, and for each sample, the BEs of the BFM parameters were obtained. The BEs from the original and five simulation datasets were then utilized to plot the FRFs and RFs of the BFM model as depicted in \figurename{s \ref{EFST_fig4}(I)} and \ref{EFST_fig4}(II). It can be witnessed that the simulated data-based curves (BFM-BM-S1 to BFM-BM-S5) for both the FRF and RF graphs have satisfactorily covered the observed data-based estimated FR and reliability curves. Consequently, one can infer that the entertained BFM model is compatible with the observed data, and none of the plotted FR and RF curves can be denied based on the predictive simulation view. 

\subsection{Illustration II: Electrical Appliance Failure and Survival Times}\label{S6.2}
\,\indent This illustration considers the AFST data \citep{Lawless2003}. The observed failures were categorized into 18 different COFs ($C_{1}$ for $i=1, \cdots, 18$). Though, out of the 33 recorded failures, only 7 COFs were represented, and among the 7, only $C_{6}$ and $C_{9}$ emerged more than twice. Thus, considering these features of the data, \cite{Lawless2003} suggested focusing on $C_{9}$ for an illustration. Hence, he redefines the cause-specific identities as $C_{1}$ if failure emerges due to $C_{9}$ and $C_{2}$ if it occurs due to any of the remaining 17 different COFs. Note that in this study, the AFSTs units are 1000 cycles to failure.

The $C_{1}$, $C_{2}$ and combined data FR characteristics are, accordingly, BFR, IFR, and RCFR curves, as portrayed in \figurename{ \ref{figTTT}}(II). 
Similar to Section \ref{S6.1_2}, we employ the BFM to model the AFST data under the two established estimation techniques and subsequently evaluate the model's performance against the several models described in \tablename{ \ref{model_tab1}}. 

\subsubsection{Posterior Estimates and Plots}\label{S6.1_2}
\,\indent In this part, we repeat the Bayesian steps discussed in Section \ref{S6_1_1}, maintaining the same number of parallel chains and iterations. 

\begin{figure}[h!]
	\centering	
	\includegraphics[width=3.3in, height=3.25in, keepaspectratio=false]{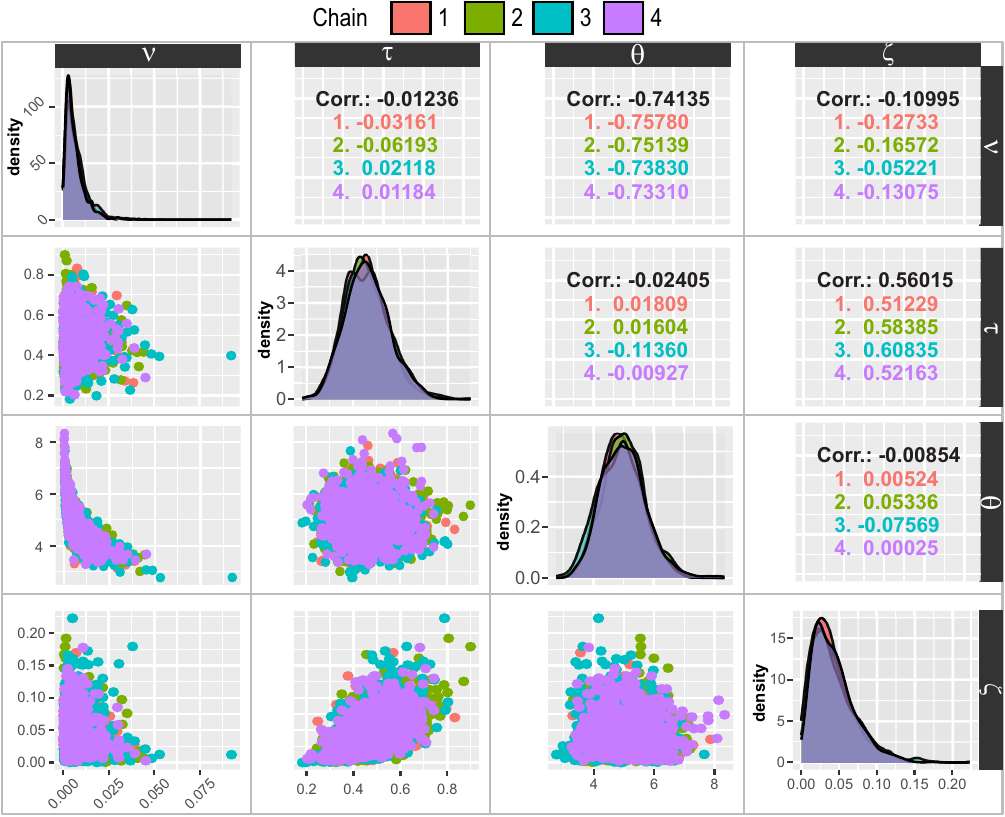}
	\centering
	\caption{\small Scatter graph depicting (\textit{up-diagonal}) posterior correlation estimates, (\textit{diagonal}) posterior density estimates, and (\textit{below-diagonal}) scatter diagram between the bi-variate of the BFM parameters posterior samples; AFST data.}
	\label{AFST_fig1}
\end{figure}

We offer the posterior diagnostic graph in \figurename{ \ref{AFST_fig1}}. The up and below diagonal of the figure shows the chain-specific and overall bi-variate posterior correlation estimates between the BFM parameter pairs, and the $\iota_{0}=4$ parallel chains scatter graphs describing the level of relations between the bi-variate posterior sample pairs, respectively. Different from the first illustration, higher and relatively higher average posterior correlation estimates of 0.74135 and 0.56015 are recorded between ($\nu$ and $\theta$) and ($\zeta$ and $\tau$), respectively, from Dhillon and exponential power components in BFM model. Similar findings are visually seen in the corresponding scatter diagrams. Nevertheless, the posterior correlations between other pairs of the parameters are observed to be very weak. For instance, the chain-wise correlation values between the BFM shape parameters ($\theta$ and $\tau$) are obtained as 0.01809, 0.01604, -0.11360, and -0.00927. These correlation values may translate to less dependence between both shape parameters from Dhillon and exponential power components. To explore the strength of the application of HMC-SA, the kernel estimates show that the posterior densities are single-modal and bell-shaped for the majority of the BFM parameters, as shown in the diagonal of \figurename{ \ref{AFST_fig1}}. In addition, the R-hat estimates of 1.0047, 1.0004, 1.0000, and 1.0000 for, accordingly, $\nu$, $\theta$, $\tau$, and $\zeta$, have clearly established the convergence of the HMC-SA-based samples.

Using the obtained posterior samples, we numerically determined the BEs of the BFM parameter alongside the HPDIs of the parameters. We also applied the MLM to determine the OMPs and ACIs of the BFM parameters. \tablename{ \ref{AFST_tab1}} reports the parameters' BEs, OMPs, and interval estimates. It is essential to note that the BEs and OMPs of the BFM's scale parameters ($\nu$ and $\zeta$) are the same. Whereas the BE of the $\theta$ is 0.01760 greater than its associated OMP, and $\tau$ is 0.01320 less than its OMP counterpart. The BM has recorded smaller St-Devs than the MLM for all BFM parameters. These smaller St-Devs for the BFM entail narrowed-width  HPDIs for the BEs, which is better than the ACIs. Comparing the proposed BM and MLM based on the three NPMs, as shown \tablename{ \ref{AFST_tab1}}, the MLM has shown a better strength over the BM in estimating the BFM parameters based on  NPM$_{\text{KS}}$ and NPM$_{\text{AD}}$. The MLM has, however, slightly performed below the BM based on NPM$_{\text{CvM}}$, which gives the highest $P_{\text{KS}}$-value of 0.4717 compared to 0.4694 of the MLM. The MTTF estimates under the two techniques are also portrayed in \tablename{ \ref{AFST_tab1}}.

\begin{table}[h!]
	\caption{\small  The BEs, OMPs, St-Devs, 95\% HPDIs, and ACIs of the BFM parameters for modeling AFST data. The three NPMs are provided to assess the two approaches.} 
	\label{AFST_tab1} 
	\centering
	\begin{adjustbox}{width=4.5in}
		\begin{tabular}{lcccc} \toprule 
			
			Parameter	& $	\widehat{\nu}$	&$	\widehat{\theta}$ &$	\widehat{\tau}$&$	\widehat{\zeta}$\\	\midrule
			
			Bayes (St-Dev)	&0.0054 (0.0059)&4.9648 (0.7196)&0.4569 (0.0940)&0.0419 (0.0050)\\	
			95\% HPD Interval&[0.0004, 0.0185]&[3.5940, 6.3976]&[0.2902, 0.6507]&[0.0012, 0.0987]\\
			R-hat&1.0047&1.0004&1.0000&1.0000\\	\cline{2-5}
			NPM$_{\text{KS}}(P_{\text{KS}})$&0.0872 (0.9446)&&&\\
			NPM$_{\text{AD}}(P_{\text{AD}})$&0.3874 (0.3684)&&$\widehat{\text{MTTF}}_{\text{BM}}\rightarrow$&2220.42\\
			NPM$_{\text{CVM}}(P_{\text{CVM}})$&0.0520 (0.4717)&&&\\
			\cline{2-5}	
			MLE(St-Dev) &0.0054 (0.00751)&4.9472 (1.2145)&0.4701 (0.1522)&0.0419 (0.0459)\\		
			95\% AC Interval&[0.0009,  0.0200]&[2.5670, 7.3275]&[0.1717, 0.7684]&[0.0000, 0.1319]\\	\cline{2-5}
			NPM$_{\text{KS}}(P_{\text{KS}})$&0.0785 (0.9773)&&&\\
			NPM$_{\text{AD}}(P_{\text{AD}})$&0.3857 (0.3718)&&$\widehat{\text{MTTF}}_{\text{MLM}}\rightarrow$&2254.41\\
			NPM$_{\text{CVM}}(P_{\text{CVM}})$&0.0521 (0.4694)&&&\\ \bottomrule
		\end{tabular}
	\end{adjustbox}
\end{table}

\subsubsection{BFM Performance Against Some Recent CR Models}
\,\indent
In this segment, we study the robustness of the BFM model using MLM against the five candidates given in \tablename{ \ref{model_tab1}} for the analysis of AFST data. The models' OMPs are shown in \tablename{ \ref{AFST_tab2}}. For evaluating the BFM modeling performance against the five competing candidates, the OMPs, St-Devs, and the seven metrics are reported in \tablename{ \ref{AFST_tab2}}. Furthermore, we determined the ranks of the metrics. The ranks' average is also given in the last column of \tablename{ \ref{AFST_tab2}}

The seven metric results established that the trained BFM model is a better option for modeling the AFST data. This selection corresponds to the BFM given the minimum values from all the seven metrics as boldly marked in \tablename{ \ref{AFST_tab2}}. Similarly, the rank-wise evaluation of these metrics has revealed the BFM leading position, as reported in the tenth column of \tablename{ \ref{AFST_tab2}}. We notice how the PM$_{\text{BC}}$ metric, which combined the strength of PM$_{\text{AIC}}$ and PM$_{\text{BIC}}$ metrics, produces the least value for the proposed BFM. Thus, the BFM proves its robustness over the APD, FACG, FAEPG, EAddW, and GExtEW models for modeling the FTs of electrical appliances. In the present illustration, the rank-wise evaluations have shown a similar performance between APD and FACG on average after the BFM. It is also noted that the trained GExtEW and FAEPG models have the least accurate description of the data, as depicted by the average ranks.  

\tablename{ \ref{AFST_tab3}} provides the estimated probabilities of failure ($\widehat{F}_{C_{1}}$ and $\widehat{F}_{C_{2}}$) caused by $C_{1}$ and $C_{2}$ for the trained BFM, APD, FACG, and FAEPG models. The table also reported the empirical probabilities of failure, which are 0.515 and 0.485, respectively, due to $C_{1}$ and $C_{2}$. The estimated probabilities from the trained BFM (BFM-BM) have attributed 62.6\% to $C_{1}$ and  37.4\%  to $C_{2}$, which are also much closer to the empirical values of 51.5\% and 48.5\% compare to the (63.5\%, 36.5\%), (90.0\%, 10.0\%), (99.5\%, 0.50\%), and (99.8.0\%, 0.20\%), respectively, from the BFM-MLM, APD, FACG and FAEPG.

\begin{table}[h!]
	\caption{\small The OMPs with St-Devs within parenthesis and performance metrics of BFM and other competing candidates for AFST data fitting.} 
	\label{AFST_tab2} 
	\centering
	\begin{adjustbox}{width=4.5in}
		\begin{tabular}{cccccccccc} \toprule 
			\large
			\underline{	Model}	&\underline{ML estimate(St-Dev)} &\multicolumn{7}{c}{\underline{Statistics}}&	\\  
			$\downarrow$		&$\downarrow$&PM$_{-\text{log}\underline{L}}$ &PM$_{\text{AIC}}$&PM$_{\text{BIC}}$&PM$_{\text{BC}}$&NPM$_{\text{KS}}$&NPM$_{\text{AD}}$&NPM$_{\text{CVM}}$&$\alpha_{R}$	\\ 		
			&& &&&&$(P_{\text{KS}})$&$(P_{\text{AD}})$&$(P_{\text{CVM}})$&\\	\cline{3-8}
			BFM	& $\widehat{\nu}=5.4e^{-3}(7.5e^{-3}),   $&\textbf{53.693}$\color{red}{^{1}}$&\textbf{115.39}$\color{red}{^{1}}$&\textbf{121.37}$\color{red}{^{1}}$&\textbf{128.82}$\color{red}{^{1}}$&\textbf{0.078}$\color{red}{^{1}}$&\textbf{0.386}$\color{red}{^{1}}$&\textbf{0.052}$\color{red}{^{1}}$&$\color{red}{1}$	\\ 
			&$\widehat{\tau}=0.470(0.152),  $&&&&&(0.977)&(0.372)&(0.469)&\\ 
			& $\widehat{\theta}=4.947(1.214), $&&&&&&&&\\
			&$\widehat{\zeta}=0.042(0.046) $&&&&&&&&\\
			APD	&$\widehat{\alpha}=0.108(0.093),  $ &55.192$\color{red}{^{3}}$&118.38$\color{red}{^{3}}$&124.37$\color{red}{^{3}}$&131.82$\color{red}{^{3}}$&0.116$\color{red}{^{2}}$&0.629$\color{red}{^{2}}$&0.104$\color{red}{^{2}}$	&$\color{red}{\approx3}$\\ 
			&$\widehat{\beta}=0.109(0.077), $ 
			&&&&&(0.719)&(0.092)&(0.095)&\\
			&$\widehat{\lambda}=1.016(0.244),  $ 
			&&&&&&&&\\
			&$ \widehat{\theta}=17.76(12.97)  $ 
			&&&&&&&&\\
			
			FACG	&$\widehat{\gamma}=0.581(0.054),,  $ &54.962$\color{red}{^{2}}$&117.92$\color{red}{^{2}}$&123.91$\color{red}{^{2}}$&131.36$\color{red}{^{2}}$&0.155$\color{red}{^{5}}$&1.110$\color{red}{^{5}}$&0.205$\color{red}{^{5}}$&	$\color{red}{\approx3}$\\ 
			&$ \widehat{\alpha}=0.202(0.053),$ &&&&&(0.365)&(0.0067)&(0.004)&	\\ 
			&$ \widehat{\lambda}=90.46(1.596),  $ &&&&&&&&	\\ 
			&$  \widehat{\theta}=709.8(12.48) $ &&&&&&&&	\\ 			
			FAEPG	&$\widehat{\alpha}=0.795(0.116),  $ &55.337$\color{red}{^{4}}$&118.67$\color{red}{^{4}}$&124.66$\color{red}{^{4}}$&132.11$\color{red}{^{4}}$&0.191$\color{red}{^{6}}$&1.274$\color{red}{^{6}}$&0.246$\color{red}{^{6}}$	&$\color{red}{\approx5}$\\ 
			&$\widehat{\gamma}=0.248(0.034),$ 
			&&&&&(0.157)&(0.002)&(0.001)&\\
			&$\widehat{\lambda}=88.92(1.435),   $ 
			&&&&&&&&\\
			&$ \widehat{\theta}=697.6(11.22)  $ 
			&&&&&&&&\\
			EaddW&$\widehat{\alpha}=0.241(0.147),  $ &56.242$\color{red}{^{5}}$&122.48$\color{red}{^{5}}$&129.978$\color{red}{^{5}}$&135.98$\color{red}{^{5}}$&0.136$\color{red}{^{3}}$&0.778$\color{red}{^{3}}$&0.136$\color{red}{^{3}}$	&$\color{red}{\approx4}$\\ 
			&$  \widehat{\beta}=1.540(0.337), $ 
			&&&&&(0.532)&(0.039)&(0.035)&\\
			&$\widehat{\theta}=74.79(30.22)$&&&&&&&&\\
			&$\widehat{\gamma}=3.755(0.480),$&&&&&&&&\\
			&$\widehat{\lambda}=0.052(0.023),$&&&&&&&&\\
			&$\widehat{\theta}=74.79(30.22)$&&&&&&&&\\
			GExtEW&$\widehat{\alpha}=1.3e^{3}(11.98),  $ &57.097$\color{red}{^{6}}$&124.19$\color{red}{^{6}}$&131.68$\color{red}{^{6}}$&137.68$\color{red}{^{6}}$&0.151$\color{red}{^{4}}$&0.898$\color{red}{^{4}}$&0.169$\color{red}{^{4}}$	&$\color{red}{\approx5}$\\ 
			&$ \widehat{\beta}=1.071(5.7e^{-3}),  $ 
			&&&&&(0.401)&(0.019)&(0.0124)&\\
			&$\widehat{\gamma}=8.3e^{-4}(4.1e^{-4}),$&&&&&&&&\\
			&$\widehat{\lambda}=3.0e^{-3}(6.4e^{-4}),$&&&&&&&&\\
			&$\widehat{c}=26.97(2.026)$&&&&&&&&\\
			\bottomrule
		\end{tabular}
	\end{adjustbox}
\end{table}	

\begin{table}[h!]
	\caption{\small Empirical and estimated probabilities of failures due to $C_{1}$ and $C_{2}$ for the BFM and other competing models; AFST data} 
	\label{AFST_tab3} 
	\centering
	\begin{adjustbox}{width=3.3in}
		\begin{tabular}{lcccccc} \toprule 
			&Empirical&BFM-BM &BFM-MLM&APD&FACG&FAEPG\\
			\cline{2-7}
			$	\widehat{F}_{C_{1}}$	&0.515&0.626&0.635&0.900&0.995&0.988	\\
			
			$	\widehat{F}_{C_{2}}$	 &0.485&0.374&0.365&0.100&0.005&0.012	\\
			
			\bottomrule
		\end{tabular}
	\end{adjustbox}
\end{table}
\subsubsection{Model Compatibility}
\begin{figure}[h!]
	\centering	
	\includegraphics[width=4.5in, keepaspectratio=false]{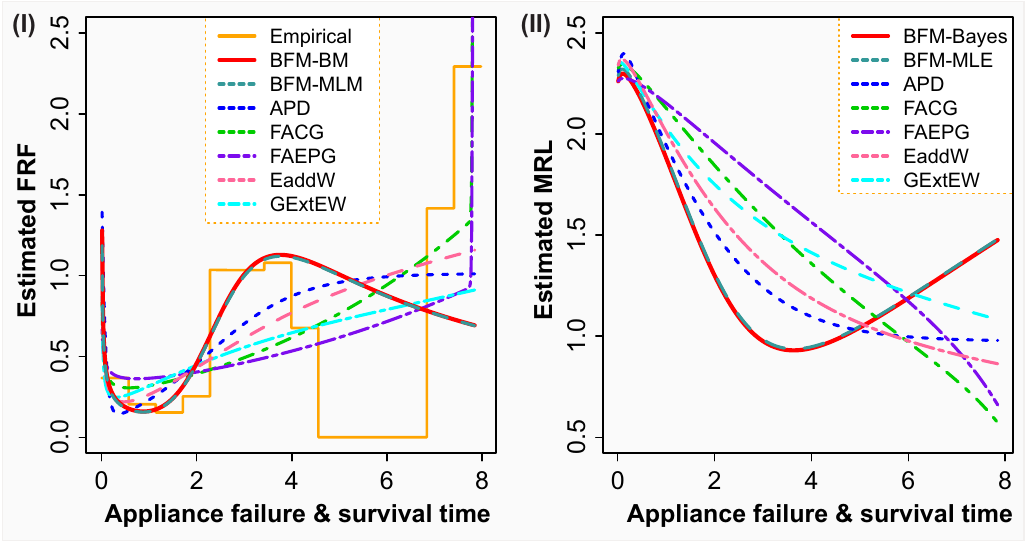}
	\centering
	\caption{\small (I) Empirical and estimated FRFs and (II) estimated MRLFs of BFM and other candidates, portraying the reciprocal relationship between the two reliability features on modeling AFST data.}
	\label{AFST_fig2}
\end{figure}

\,\indent \figurename{s \ref{AFST_fig2}(I)} and \ref{AFST_fig2}(II) depict the FRFs $r(m)$ and MRLFs $\mu(\tilde{x})$, respectively, of the trained models on AFST data, to demonstrate the compatibility of the proposed BFM against the five competing candidates.  \figurename{s \ref{AFST_fig2}(I)} also shows the empirical FR of the data. The FR curve estimates of the proposed BFM model under the BM and MLM have fundamentally the same features and have, to some certain level, mimicked the empirical FR of the data. Nevertheless, The BFM has fallen short of correctly identifying the last part of the AFST data's empirical RCFR. As shown by the TTT-transform graph in \figurename{ \ref{figTTT}(II)}, the data has a four-segment FR curve. The proposed BFM model was able to identify the first three portions but failed to estimate the last IFR  at the wear-out period. Notwithstanding, the BFM shows the best FR prediction in \figurename{ \ref{AFST_fig2}(I)} among the competing candidates. Besides that, the proposed model has managed to predict the reciprocal relationship between its trained FRF and MRLF as shown in  \figurename{s \ref{AFST_fig2}(I)} and \ref{AFST_fig2}(II). 

\begin{figure}[h!]
	\centering	
	\includegraphics[width=4.5in, height=3.2in, keepaspectratio=false]{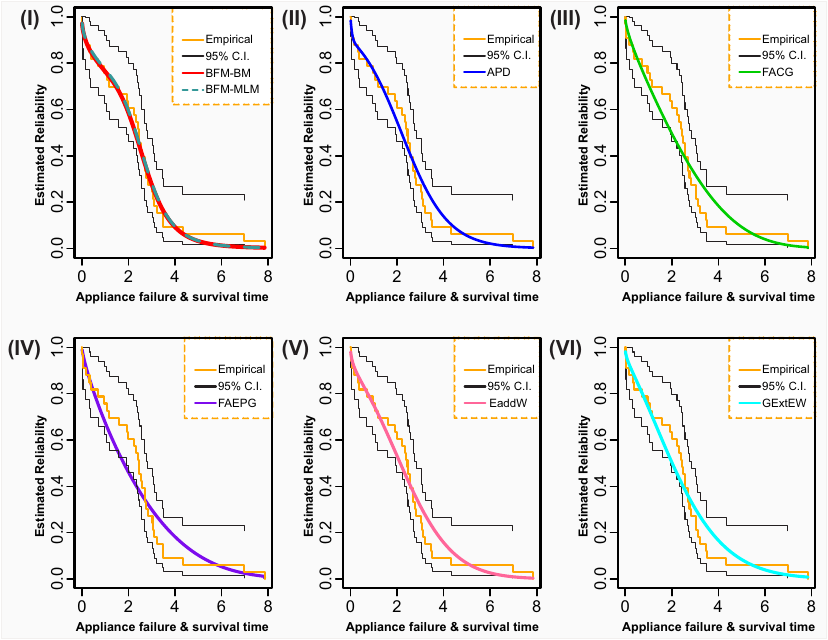}
	\centering
	\caption{\small Empirical and reliability estimates of BFM and five competing models; AFST data.}
	\label{AFST_fig3}
\end{figure}

Besides visualizing the FRFs and MRLFs of the trained models, we plotted the empirical and fitted RFs of these models in  \figurename{s \ref{AFST_fig3}(I)}--\ref{AFST_fig3}(VI). It is undeniable to see that the reliability curves of the BFM model (as shown in \figurename{ \ref{AFST_fig3}(I)} for both BM and MLM) follow the empirical distribution much closer than the other competing models. Arguably, the excellent performance of the BFM, both numerically and visually, may not be unrelated to the flexibility contributions given by the Dhillon and exponential power components of the model. Although the BFM presents the best description of this case study, it is additionally crucial in future work to explore how the RCFR curve can be predicted accurately by the model since it possesses the feature as shown in \figurename{ \ref{F1}(III)}.


\begin{figure}[h!]
	\centering	
	\includegraphics[width=3.5in, keepaspectratio=false]{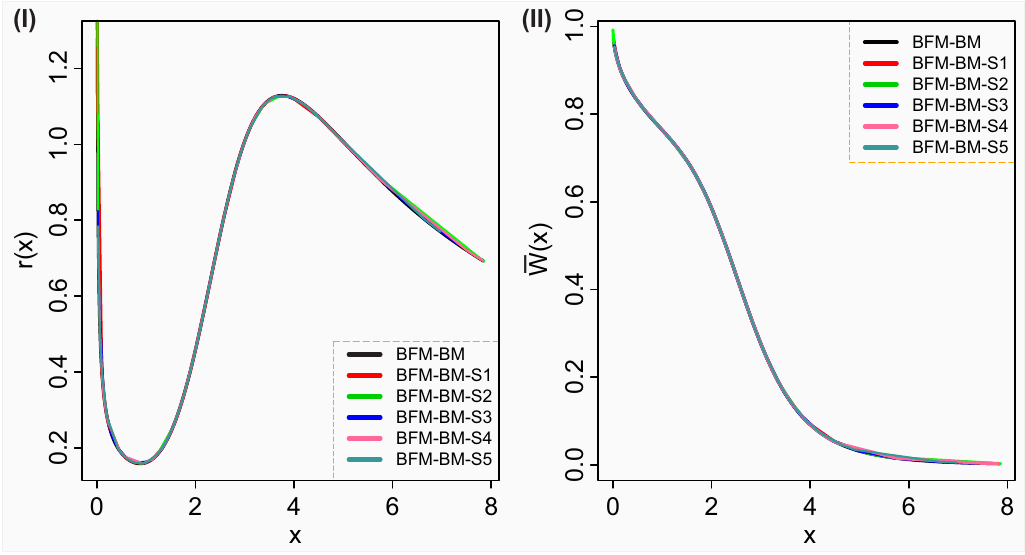}
	\centering
	\caption{\small Empirical and reliability estimates of BFM and five competing models; AFST data.}
	\label{AFST_fig4}
\end{figure}

To additionally analyze the compatibility of the proposed BFM for the Bayesian modeling of the AFST data, five different random samples of equal size with the observed data were drawn from the trained BFM model, and for each sample, the BEs of the BFM parameters were obtained. Given that our focus is on statistical reliability, the six sets of BEs from the observed and simulation datasets were used to plot the FRF and RF of the BFM model as depicted in \figurename{s \ref{AFST_fig4}(I)} and \ref{AFST_fig4}(II). It can be noticed that the simulated data-based curves (BFM-BM-S1 to BFM-BM-S5) for both the FRF and RF graphs have nicely covered the observed data-based estimated FR and reliability curves. Hence, one can deduce that the BFM model is compatible with the observed data, and none of the five FRF and RF graphs can be rejected based on the idea of the predictive simulation. 

\section{Conclusions}\label{S6}
\,\indent We introduce a minimum failures-based competing risk (CR) model named the Bi-Failure Modes (BFM) model. The BFM model is constructed such that the distribution of FTs due to one failure type follows a Dhillon model, and that of the other cause is attributed to an exponential power model. The proposed BFM model possesses the BFR, IBBFR, and RCFR characteristics suitable for modeling varied CR datasets, assuming independence among the failure types. A detailed study of several reliability attributes of the BFM model is presented, including the MRLF, MTTF expression, and cause-specific failure probabilities. We also investigate the fundamental reciprocal relationships between the MRLF and the failure rate function (FRF). We propose the HMC-SA-based Bayesian method for estimating the BFM model parameters and its reliability attributes to offer greater computational efficiency and faster inference. Two CR datasets characterized by BFR and RCFR behaviors are employed to demonstrate the BFM adequacy in the CR data study. The BFM has portrayed its adequate potential for modeling the two datasets under the BM. The recently introduced Bridge Criterion metric and six other metrics were deplored to evaluate the BFM modeling accuracy against five recent methodologies under the maximum likelihood technique. In this case, the trained BFM model is far more compatible with the two datasets than the five competing candidates.

Despite its potential over several competing models, the BFM has fallen short of correctly identifying the last part of the empirical RCFR under illustration II. It is, therefore, very crucial in future work to explore how the RCFR curve can be predicted accurately by the BFM model since it possesses the feature as shown in \figurename{ \ref{F1}(III)}. As pointed out in Section \ref{S1}, the BFM model is mainly built for CR and complex data studies, where the combined FTs have non-monotone FR behaviors. It can, however, be tested on monotone FR data in future studies, as usually recommended by experts.


\bibliography{submission}
\bibliographystyle{apa}

\end{document}